\documentclass[acmsmall,usenames,dvipsnames]{acmart}

\AtBeginDocument{%
  }

\setcopyright{acmlicensed}
\copyrightyear{XXXX}
\acmYear{XXXX}
\acmDOI{XXXXXXX.XXXXXXX}

\usepackage{enumitem}

\usepackage{listings}

\usepackage{xspace}

\usepackage{pifont}
\usepackage{wasysym}

\usepackage{tabularx}
\usepackage{booktabs}
\usepackage{makecell}
\usepackage{multirow}
\usepackage{longtable}

\usepackage{microtype}
\microtypecontext{spacing=nonfrench}

\newcolumntype{C}[1]{>{\centering\arraybackslash}p{#1}}

\definecolor{pblue}{rgb}{0.13,0.13,1}
\definecolor{pgreen}{rgb}{0,0.5,0}
\definecolor{pred}{rgb}{0.9,0,0}
\definecolor{pgrey}{rgb}{0.46,0.45,0.48}
\definecolor{backcolour}{rgb}{0.95,0.95,0.92}
\definecolor{Gray}{gray}{0.85}

\lstdefinestyle{mystyle}{
  language=Java,
  showspaces=false,
  showtabs=false,
  breaklines=true,
  showstringspaces=false,
  breakatwhitespace=true,
  captionpos=b,
  basicstyle=\scriptsize\ttfamily,
  moredelim=[il][\textcolor{pgrey}]{$$},
  moredelim=[is][\textcolor{pgrey}]{\%\%}{\%\%}
}

\newcounter{dgcounter}
\newcommand{\newdg}[1]{\noindent\refstepcounter{dgcounter}(\textbf{DG\arabic{dgcounter})}\label{#1}}
\newcommand{\dgref}[1]{\textbf{DG\ref{#1}}}

\makeatletter
\newcommand\flawtag[2]{#1\def\@currentlabel{#1}\label{#2}}
\makeatother

\begin{document}
\newcommand{\tool}{{\small MASC}\xspace}
\newcommand{\tools}{{\small MASC's}\xspace}
\newcommand{\mascweb}{{\small{\tool{} Web}\xspace}}

\newcommand{\mdroid}{{\small MDroid$+$}\xspace}
\newcommand{\crashscope}{{\small CrashScope}\xspace}
\newcommand{\mse}{$\mu$SE\xspace}

\newcommand{\detector}{crypto-detector\xspace}
\newcommand{\Detector}{Crypto-detector\xspace}
\newcommand{\detectors}{crypto-detectors\xspace}
\newcommand{\Detectors}{Crypto-detectors\xspace}
\newcommand{\oakland}{S\textit{\&}P'22~\cite{ACK+22}\xspace}
\newcommand{\cipher}{\texttt{Cipher}\xspace}
\newcommand{\ciphergetinstance}{\texttt{\small Cipher.\-get\-Instance\-(<parameter>)}\xspace}
\newcommand{\messageDigestInstance}{\texttt{\small Message\-Digest.\-get\-Instance\-(<parameter>)}\xspace}
\newcommand{\messageDigest}{\texttt{\small MessageDigest}\xspace}
\newcommand{\des}{\texttt{des}\xspace}
\newcommand{\DES}{\texttt{DES}\xspace}
\newcommand{\AES}{\texttt{AES}\xspace}
\newcommand{\sarif}{\texttt{SARIF}\xspace}
\newcommand{\ECB}{\texttt{ECB}\xspace}
\newcommand{\MDFIVE}{\texttt{MD5}\xspace}
\newcommand{\muse}{$\mu$SE\xspace}

\newcommand{\trustManager}{\texttt{\small TrustManager}\xspace}
\newcommand{\xtrustManager}{\texttt{\small X509Trust\-Manager}\xspace}
\newcommand{\extendedxtrustManager}{\texttt{\small X509ExtendedTrustManager}\xspace}
\newcommand{\getAcceptedIssuers}{\texttt{\small getAcceptedIssuers}\xspace}
\newcommand{\checkServerTrusted}{\texttt{\small checkServerTrusted}\xspace}
\newcommand{\checkClientTrusted}{\texttt{\small checkClientTrusted}\xspace}

\newcommand{\certificateException}{\texttt{\small CertificateException}\xspace}

\newcommand{\hostnameVerifier}{\texttt{\small HostnameVerifier}\xspace}
\newcommand{\trycatch}{\texttt{\small try-catch}\xspace}

\newcommand{\ivparameterspec}{\texttt{IvParameterSpec}\xspace}

\newcommand{\cryptoguard}{CryptoGuard\xspace}
\newcommand{\cryptoguardmultidex}{{$2,709$}\xspace}
\newcommand{\cryptoguarddownload}{{$4,353$}\xspace}
\newcommand{\cryptoguardtotal}{{$6,181$}\xspace}
\newcommand{\cryptoguardpercent}{{$62.23$\%}\xspace}
\newcommand{\cryptoguardandroiddottotal}{{$673$}\xspace}
\newcommand{\cryptoguardandroiddot}{{$383$}\xspace}
\newcommand{\cryptoguardendswithandroid}{{$290$}\xspace}
\newcommand{\cryptoguardandroiddottotalpercent}{{$10.89$\%}\xspace}
\newcommand{\androiddot}{{\texttt{android.}}\xspace}

\newcommand{\totaloperators}{{$19$}\xspace}

\newcommand{\consistencycheck}{}
\newcommand{\crysl}{CrySL\xspace}
\newcommand{\cognicrypt}{CogniCrypt\xspace}
\newcommand{\xanitizer}{Xanitizer\xspace}
\newcommand{\coverity}{Tool$_X$\xspace}
\newcommand{\spotbug}{SpotBugs\xspace}
\newcommand{\spotbugfull}{SpotBugs with FindSecBugs\xspace}
\newcommand{\qark}{QARK\xspace}
\newcommand{\qpid}{Apache Qpid Broker-J\xspace}
\newcommand{\shiftleft}{ShiftLeft\xspace}
\newcommand{\codeqlgcs}{Github Code Security\xspace}
\newcommand{\codeqllgtm}{LGTM\xspace}
\newcommand{\sonarqube}{SonarQube\xspace}

\newcommand{\codeguru}{Amazon CodeGuru Security\xspace}
\newcommand{\snyk}{Snyk\xspace}
\newcommand{\codiga}{Codiga\xspace}
\newcommand{\deepsource}{DeepSource\xspace}
\newcommand{\cryptoguardupdate}{CryptoGuard version 04.05.03\xspace}
\newcommand{\cognicryptupdate}{CogniCrypt version 2.8.0\xspace}
\newcommand{\spotbugsupdate}{SpotBugs 4.0.4 with FindSecBugs version 1.12.0 \xspace}
\newcommand{\coverityupdate}{Coverity version 2022.6.0 \xspace}
\newcommand{\sonarqubeupdate}{SonarQube Community Edition 9.9.0.65466\xspace}
\newcommand{\qarkupdate}{QARK version 4.0.0 \xspace}
\newcommand{\shiftleftupdate}{ShiftLeft version 2.1.1\xspace}
\newcommand{\codeqlversion}{GitHub Code Security CI/CD Apr 2024\xspace}
\newcommand{\newlgtm}{GitHub Code Scan v2.12.6 with LGTM Ruleset\xspace}

\newcommand{\coverityshort}{TX\xspace}
\newcommand{\xanitizershort}{XT\xspace}
\newcommand{\cryptoguardshort}{CG\xspace}
\newcommand{\shiftleftshort}{SL\xspace}
\newcommand{\qashort}{QA\xspace}
\newcommand{\codeqlgcsshort}{GCS\xspace}
\newcommand{\codeqllgtmshort}{LGTM\xspace}
\newcommand{\cryslshort}{CL\xspace}
\newcommand{\sportbugsshort}{SB\xspace}
\newcommand{\cognicryptshort}{CC\xspace}

\newcommand{\codegurushort}{AC\xspace}
\newcommand{\sonarqubeshort}{SQ\xspace}
\newcommand{\snykshort}{SY\xspace}
\newcommand{\codigashort}{CD\xspace}
\newcommand{\deepsourceshort}{DS\xspace}
\newcommand{\newcryptoguardshort}{nCG\xspace}
\newcommand{\newcognicryptshort}{nCC\xspace}
\newcommand{\newspotbugsshort}{nSB\xspace}
\newcommand{\newcoverityshort}{nTX\xspace}
\newcommand{\newqarkshort}{nQA\xspace}
\newcommand{\newshiftleftshort}{nSL\xspace}
\newcommand{\newcqversion}{nLG\xspace}
\newcommand{\newlgtmshort}{nGC\xspace}

\newcommand{\countflaws}{{$19$}\xspace}
\newcommand{\countFlawClasses}{{$5$}\xspace}
\newcommand{\countFlawClassesText}{{five}\xspace}

\newcommand{\no}{\remove{\ding{55}}}
\newcommand{\ye}{\ding{51}}
\newcommand{\nye}{\ding{51}\ding{51}}
\newcommand{\pa}{\LEFTcircle}
\newcommand{\pr}{\LEFTcircle}
\newcommand{\na}{\texttt{-}}
\newcommand{\nl}{\text{\O}}
\newcommand{\np}{\text{\O}}

\newcommand{\fcincomplete}{Flaw Class 1 (FC1): Incomplete Analysis of Target Code\xspace}
\newcommand{\fcdifferentcase}{Flaw Class 1 (FC1): String Case Mishandling\xspace}
\newcommand{\fcvalueresoluion}{Flaw Class 2 (FC2): Incorrect Value Resolution\xspace}
\newcommand{\fcomplexinheritance}{Flaw Class 3 (FC3): Incorrect Resolution of Complex Inheritance and Anonymous Objects\xspace}
\newcommand{\fcgenericnoise}{Flaw Class 4 (FC4): Insufficient Analysis of Generic Conditions in Extensible Crypto-APIs\xspace}%
\newcommand{\fcspecificnoise}{Flaw Class 5 (FC5): Insufficient Analysis of Context-specific, Conditions in Extensible Crypto-APIs\xspace}

\newcommand{\totalMutantApplications}{{$17$}\xspace}
\newcommand{\totalMutations}{{$20,303$}\xspace}
\newcommand{\totalMutationsAndroid}{{$2,515$}\xspace}
\newcommand{\totalMutationsJava}{{$17,788$}\xspace}
\newcommand{\totalReachMutant}{{$20,165$}\xspace}
\newcommand{\testedAndroidApps}{{$13$}\xspace}
\newcommand{\testedJavaComponents}{{$4$}\xspace}
\newcommand{\testedCryptoApps}{{$7$}\xspace}
\newcommand{\totalToolsUsed}{{$9$}\xspace}
\newcommand{\totalToolsUsedText}{{nine}\xspace}

\newcommand{\newMutantApplications}{{$15$}\xspace}
\newcommand{\newTotalMutantApplications}{{$32$}\xspace}
\newcommand{\newMutations}{{$30,236$}\xspace}
\newcommand{\newTotalMutations}{{$50,539$}\xspace}

\newcommand{\newTotalToolsUsed}{{$12$}\xspace}

\newcommand{\mainScope}{main\xspace}
\newcommand{\similarityScope}{similarity scope\xspace}
\newcommand{\exhaustiveScope}{exhaustive scope\xspace}
\newcommand{\totalMisuseCases}{{107}\xspace}
\newcommand{\newMisuseCases}{{4}\xspace}
\newcommand{\newMisuseSources}{{19}\xspace}
\newcommand{\implementedMisuseCases}{{$22$}\xspace}

\newcommand{\dexlib}{{\texttt{dexlib2}}\xspace}

\newcommand{\projtodo}[1]{{\color{red}{\bf TODO:}#1}}

\newcommand*{\img}[1]{%
    \raisebox{-.2\baselineskip}{%
        \includegraphics[
        height=0.85\baselineskip,
        width=0.85\baselineskip,
        keepaspectratio,
        ]{#1}%
    }%
}

\newboolean{showcomments}

\setboolean{showcomments}{true}

\ifthenelse{\boolean{showcomments}}
  {\newcommand{\nb}[2]{
    \fbox{\bfseries\sffamily\scriptsize#1}
    {\sf\small$\blacktriangleright$\textit{#2}$\blacktriangleleft$}
   }
   \newcommand{\cvsversion}{\emph{\scriptsize$-$Id: macro.tex,v 1.9 2005/12/09 22:38:33 giulio Exp $}}
  }
  {\newcommand{\nb}[2]{}
   \newcommand{\cvsversion}{}
  }

\newcommand\myparagraph[1]{\noindent\underline{\bf {#1}:}}
\newcommand\myparagraphnew[1]{\noindent{\bf {#1}:}}
\newcommand\emparagraph[1]{\noindent {\em {#1}:}}
\newcommand\budget[1]{{\color{red}\myparagraph{Budget}{#1} pages}}
\newcommand{\bnum}[1]{{(#1)}}

\newcommand{\addnewline}{\\}

\newcommand{\cancel}[1]{{\leavevmode\color{RubineRed}{\sout{\xspace#1}}}}
\newcommand{\edit}[2]{{\leavevmode\color{RubineRed}{\sout{#1}}}{\color{blue}{\xspace#2}}}
\newcommand{\editl}[2]{{\leavevmode\color{RubineRed}{\addnewline\sout{#1}}}{\color{blue}{\addnewline\xspace#2\addnewline}}}
\newcommand{\rewrite}[2]{{\leavevmode\color{RubineRed}{\sout{#1}}}{\color{Green}{\arrow\xspace#2}}}

\newcommand{\add}[1]{{\leavevmode\color{blue}{#1}}}
\newcommand{\addamit}[1]{{\leavevmode\color{blue}{#1}}}
\newcommand{\addnew}[1]{{\leavevmode\color{blue}{#1}}}

\newcommand{\addl}[1]{{\leavevmode\color{Green}{\addnewline+ \xspace#1\addnewline}}}
\newcommand{\remove}[1]{{\leavevmode\color{red}{\xspace#1}}}
\newcommand{\removel}[1]{{\leavevmode\color{red}{\addnewline- \xspace#1\addnewline}}}

\newcommand{\grayme}[1]{{\leavevmode\color{gray}{\xspace#1}}}

\newcommand{\recycler}{\textsf{RecyclerView}\xspace}

\newcommand\finding[1]{\vspace{0.25em}\noindent\textsf{\bf Finding {#1}.}}
\newcommand\fnumber[1]{{\bf F{#1}}}
\newcommand\operator[2]{{\bf OP$_{#1}$: {\em {#2}} -- }}
\newcommand\opnumber[1]{{{\bf OP}$_{#1}$}}
\newcommand\opnumbernormal[1]{{{OP}$_{#1}$}}
\newcommand\misusenumber[1]{{{\bf M}$_{#1}$}}

\newcommand{\arrow}{{$\rightarrow$}\xspace}
\newcommand\inline[1]{{\lstinline[keywordstyle=\color{black},basicstyle=\scriptsize\ttfamily,stringstyle=\color{black}]{#1}}}
\newcommand\inlinesmall[1]{{\lstinline[keywordstyle=\color{black},basicstyle=\small\ttfamily,stringstyle=\color{black}]{#1}}}
\newcommand{\boxme}[1]{{
\begin{tcolorbox}[enhanced,skin=enhancedmiddle,borderline={1mm}{0mm}{MidnightBlue}]
    \textbf{Insight: } #1 \end{tcolorbox}
}}

\newcommand\fix[1]{{\color{blue} \nb{FIX THIS}{#1}}}
\newcommand\blue[1]{{\color{blue}{#1}}}
\newcommand{\here}{{\color{blue} \nb{***}{CONTINUE HERE}}}
\newcommand{\REF}{{\color{red} \textbf{[REFS]}}\xspace}
\newcommand{\xy}{{\color{red} \textbf{XY}}\xspace}
\newcommand\tops[1]{{\color{blue}{#1}}}
\newcommand\alert[1]{{\color{red}{#1}}}

\newcommand{\target}{\textit{target tool}\xspace}
\newcommand{\targets}{\textit{target tools}\xspace}
\newcommand{\behavior}{\textit{target behavior}\xspace}
\newcommand{\ie}{\textit{i.e.,}\xspace}
\newcommand{\eg}{\textit{e.g.,}\xspace}
\newcommand{\etc}{\textit{etc.}\xspace}
\newcommand{\etal}{\textit{et al.}\xspace}
\newcommand{\etals}{\textit{et al.'s}\xspace}
\newcommand{\aka}{\textit{a.k.a.}\xspace}

\newcommand{\mplus}{{\sc MDroid+}\xspace}

\newcommand{\secref}[1]{\S\ref{#1}\xspace}
\newcommand{\figref}[1]{Fig.~\ref{#1}\xspace}
\newcommand{\tabref}[1]{Table~\ref{#1}\xspace}
\newcommand{\Phase}{{\sc Phase}\xspace}
\newcommand{\Phases}{{\sc Phase's~}\xspace}

\newcommand{\emphquote}[1]{{\emph{`#1'}}\xspace}
\newcommand{\emphdblquote}[1]{{\emph{``#1''}}\xspace}

\newcommand{\emphbrack}[1]{\emph{[#1]}\xspace}

\newcommand{\subj}{\emphbrack{subject}}
\newcommand{\act}{\emphbrack{action}}
\newcommand{\obj}{\emphbrack{object}}
\newcommand{\prep}{\emphbrack{preposition}}
\newcommand{\objtwo}{\emphbrack{object2}}

\newcommand*\ciclednum[1]{\raisebox{.5pt}{\textcircled{\raisebox{-.9pt}
{#1}}}}

\newcommand\codel[1]{\begin{verbatim}{#1}\end{verbatim}}

\newcommand\tfootnotesymbol[1]{$^{\mathrm{#1}}$}
\newcommand\tfootnotecontent[2]{$^{\mathrm{#1}}$#2}

\title{Mutation-based Evaluation of Cryptographic API Misuse Detectors}

\author{Amit Seal Ami}\email{amitsealami@usf.edu}
\orcid{0000-0002-9455-2230}
\affiliation{
  \institution{Bellini College of AI, Cybersecurity, and Computing, University of South Florida}
  \streetaddress{}
  \city{Tampa}
  \state{Florida}
  \country{USA}
  \postcode{}
}

\author{Scott Marsden}\email{smarsden@wm.edu}
\affiliation{
  \institution{Computer Science Department, William \& Mary}
  \streetaddress{}
  \city{Williamsburg}
  \state{Virginia}
  \country{USA}
  \postcode{}
}

\author{Kevin Moran}
\email{kpmoran@ucf.edu}
\orcid{0000-0001-9683-5616}
\affiliation{%
  \institution{Department of Computer Science, University of Central Florida}
  \city{Orlando}
  \state{Florida}
  \country{USA}}

\author{Denys Poshyvanyk}
\email{denys@cs.wm.edu}
\orcid{0000-0002-5626-7586}
\affiliation{
  \institution{Computer Science Department, William \& Mary}
  \streetaddress{}
  \city{Williamsburg}
  \state{Virginia}
  \country{USA}
  \postcode{}
}

\author{Adwait Nadkarni}
\orcid{0000-0001-6866-4565}
\email{apnadkarni@wm.edu}
\affiliation{
  \institution{Computer Science Department, William \& Mary}
  \streetaddress{}
  \city{Williamsburg}
  \state{Virginia}
  \country{USA}
  \postcode{}
}
\renewcommand{\shortauthors}{Amit et al.}

\begin{abstract}
The {\em correct} use of cryptography is central to ensuring data security in modern software systems.
Hence, several academic and commercial static analysis tools have been developed for detecting and mitigating crypto-API misuse.
While developers are optimistically adopting these crypto-API misuse detectors (or \detectors) in their software development cycles, this momentum must be accompanied by a {\em rigorous understanding of their effectiveness at finding crypto-API misuse in practice}.
This paper describes the \tool framework, which enables a systematic and data-driven evaluation of \detectors using mutation testing.
We ground \tool in a comprehensive view of the problem space by developing a data-driven taxonomy of existing crypto-API misuse, containing \totalMisuseCases misuse cases organized among nine semantic clusters.
We develop 19 generalizable {\em usage-based mutation operators} and three {\em mutation scopes} that can expressively instantiate thousands of compilable variants of the misuse cases for thoroughly evaluating \detectors.
Using \tool{}, in a previous study, we evaluated {\em nine} major \detectors and discovered \countflaws{} unique, undocumented flaws that severely impact the ability of \detectors to discover misuses in practice.
This paper substantially extends our \tool{} framework
and offers updated evaluation of the \detectors in our 2022 study, in addition to 5 more, major \detectors{}.
Through this work, we find 6 new, undocumented flaws, and demonstrate that these flaws affect the \detectors{} regardless of their origin; open-source community, industry, and/or research.
We conclude with a discussion on the diverse perspectives that influence the design of \detectors{} and future directions towards building security-focused \detectors{} by design.

\end{abstract}



\maketitle

\section{Introduction}\label{sec:introduction}

Effective cryptography is critical in ensuring the security of confidential data in modern software.
However, ensuring the {\em correct use} of cryptographic primitives has historically been a hard problem, whether we consider the vulnerable banking systems from Anderson's seminal work~\cite{and93}, or the widespread misuse of cryptographic APIs (\ie\ {\em crypto-APIs}) in mobile and Web apps that can lead to the compromise of confidential financial or medical data and even the integrity of IoT devices~\cite{fhm+12,rxa+19,SSG+14,TLL+19,ZWG15,ZCD+19,kmm+19}.
In response, security researchers have developed a wide array of techniques and tools for detecting crypto-API misuse~\cite{KSA+18,rxa+19,fhm+12,ebfk13,SSG+14,spotbugs_findsecbugs,XanitizerRIGSIT,CoveritySASTSoftware,sonarqube,github_security_lab,shiftleft,lgtm} that can be integrated into the software development cycle, thereby preventing vulnerabilities at the source.
These crypto-API misuse detectors, or {\em \detectors}, play a crucial role in the security of end-user software.

Crypto-detectors have been independently used by developers for decades~\cite{BBC+10}.
They are integrated into IDEs (\eg the CogniCrypt plugin for Eclipse~\cite{cognicrypteclipse}), incorporated in the internal testing suites of organizations (\eg \cryptoguard{}~\cite{rxa+19}, integrated into Oracle's testing suite~\cite{cryptoguard_oracle}), or are currently targeted for commercialization and widespread deployment~\cite{rxa+19,cryptoguard_nsf_ttp}.
In fact, several \detectors are also being formally provisioned by code hosting services as a way of allowing developers to ensure compliance with data security standards and security best-practices (\eg Github's CodeScan initiative~\cite{github_third_party_code_scanners}).
Thus, the importance of \detectors in ensuring data security in modern Web and mobile software cannot be overstated, as key stakeholders (\ie researchers, code-hosting services, app markets, and developers) are increasingly reliant on them.
However, what is concerning is that while stakeholders are optimistically adopting \detectors, {\em we know very little regarding their actual effectiveness at finding crypto-API misuse}.
That is, beyond manually-curated benchmarks, there is no approach for {\em systematically} evaluating \detectors.
This example in Listing~\ref{lst:des-example} illustrates the gravity of this problem:
\begin{lstlisting}[frame=tb,caption={\small Instantiating ``DES'' as a cipher instance.}, label={lst:des-example},language=java]
String a="DES"; Cipher c=Cipher.getInstance(a); c.init(Cipher.ENCRYPT_MODE, k, iv);
byte enc = c.DoFinal(text);
\end{lstlisting}
In this example, we define \texttt{DES} as our algorithm of choice, and instantiate it using the \inlinesmall{getInstance} API from \inlinesmall{Cipher}.
Given that \texttt{DES} is not secure, one would expect any \detector to detect this relatively straightforward misuse.
However, two very popular \detectors, \ie \coverity{}\footnote{We have anonymized this tool in the paper as requested by its developers.} (used by over 3k+ open source Java projects), and \qark~\cite{qark_linkedin_engineering_team} (promoted by LinkedIn and recommended in security testing books~\cite{KI16,HWA+17,Hsu19}), are unable to detect this trivial misuse case as we discuss later in the paper. %
Further, one might consider manually-curated benchmarks (\eg CryptoAPIBench~\cite{ary19}, or the OWASP Benchmark~\cite{OWASP_benchmark}) as practical and sufficient for evaluating \detectors to  uncover such issues.
However, given the scale and diversity of crypto protocols, APIs, and their potential misuse, benchmarks may be incomplete, incorrect, and impractical to maintain; \eg the OWASP benchmark considered using \texttt{ECB} mode with \texttt{DES} as secure until it was reported in March 2020~\cite{owasp:mislabel}.
Thus, it is imperative to address this problem
through a reliable and evolving evaluation technique that scales to the volume and diversity of crypto-API misuse.

In this paper, we describe the first systematic, data-driven framework that leverages the well-founded approach of Mutation Analysis for evaluating Static Crypto-API misuse detectors -- the \tool framework, pronounced as {\em mask}.
Stakeholders can use \tool in a manner similar to the typical use of mutation analysis in software testing: \tool\ {\em mutates} Android/Java apps by seeding them with {\em mutants}, \ie code snippets exhibiting crypto-API misuse.
These mutated apps are then analyzed with the \detector that is the target of the evaluation, resulting in mutants that are {\em undetected}, which when analyzed further reveal design or implementation-level flaws in the crypto-detector.
To enable this workflow for practical and effective evaluation of \detectors, \tool addresses three key {\bf \em research challenges}~({\bf RCs}) arising from the unique scale and complexity of the problem domain of crypto-API misuse:

\noindent{\bf RC$_1$: }\textit{Taming the Complexity of Crypto-API Misuse -} An approach that effectively evaluates \detectors must  comprehensively express (\ie test with) relevant misuse cases {\em across all existing crypto-APIs}, which is challenging as crypto-APIs are as vast as the primitives they enable.
For instance, APIs express the initialization of secure random numbers, creation of ciphers for encryption/decryption, computing message authentication codes (MACs), and higher-level abstractions such as certificate and hostname verification for SSL/TLS.%

\noindent{\bf RC$_2$: }\textit{Instantiating Realistic Misuse Case Variations -} To evaluate crypto-detectors, code {\em instances} of crypto-API misuse must be seeded into apps for analysis.
However, simply injecting misuse identified in the wild \textit{verbatim} may not lead to a robust analysis, as it does not express the variations with which developers may use such APIs. %
Strategic and expressive {\em instantiation} of misuse cases is critical for an effective evaluation, as even subtle variations may evade detection, and hence lead to the discovery of flaws (\eg passing \texttt{DES} as a variable instead of a constant in Listing~\ref{lst:des-example}).

\noindent{\bf RC$_3$: }\textit{Scaling the Analysis -} Efficiently creating and seeding large numbers of {\em compilable} mutants {\em without significant manual intervention} is critical for identifying as many flaws in \detectors as possible. Thus, the resultant framework must efficiently scale to thousands of tests (\ie mutants).

\noindent To address these research challenges, this paper makes the following major contributions in its conference version of the paper~\oakland:

\begin{itemize}

    \item {\bf Crypto-API Misuse Taxonomy:} We construct the first comprehensive taxonomy of crypto-API misuse cases ($105$ cases, grouped into nine clusters), using a data-driven process that systematically identifies, studies, and extracts misuse cases from academic and industrial sources published over the last $20$ years. The taxonomy provides a broad view of the problem space, and forms the core building block for \tools approach, enabling it to be grounded in real misuse cases observed in the wild  ({\bf RC$_1$}).

    \item {\bf {\em Crypto}-Mutation Operators and Scopes:} We contextualize mutation testing for evaluating \detectors by designing abstractions that allow us to instantiate the misuse cases from the taxonomy to create a diverse array of feasible (\ie\ {\em compilable}) mutants. We begin by formulating a threat model consisting of 3 adversary-types that represent the threat conditions that \detectors may face in practice. We then design {\em usage-based mutation operators}, \ie general operators that leverage the common usage characteristics of diverse crypto-APIs, to expressively instantiate misuse cases from the taxonomy (addresses {\bf RC$_2$}). Similarly, we also design the novel abstraction of {\em mutation scopes} for seeding mutants of variable fidelity to realistic API-use and threats.
    \item {\bf The \tool Framework:} We implement the \tool framework for evaluating Java-based \detectors, including \totaloperators{} mutation operators that can express a majority of the cases in our taxonomy, and 3 mutation scopes. We implement the underlying static analysis to automatically instantiate thousands of {\em compilable} mutants, with manual effort limited to configuring the mutation operators with values signifying the misuse ({\bf RC$_3$}).
    \item {\bf Empirical Evaluation of Crypto-Detectors:} We evaluate $9$ major \detectors using \totalMutations{} mutants generated by \tool, and reveal \countflaws{} previously unknown flaws (several of which are design-level). A majority of these discoveries of flaws in individual detectors (\ie 45/76 or 59.2\%) are due to mutation (vs. being unable to detect the base/verbatim instantiations of the misuse case). Through the study of open source apps, we {\em demonstrate that the flaws uncovered by \tool are serious and would impact real systems}. Finally, we disclose our findings to the designers/maintainers of the affected \detectors, and further leverage these communication channels to obtain {\em their perspectives} on the flaws. These perspectives allow us to present a balanced discussion on the factors influencing the current design and testing of \detectors, as well as a path forward towards more robust tool.
\end{itemize}
This study substantially extends upon the previous work, described as follows:
\begin{itemize}
    \item {\textbf{Updating the Taxonomy}:} By applying the data-driven approach from the \oakland{} paper for recent crypto-API misuse reported in industry and academic sources from the year $2019$ to $2022$, we have extended the crypto-API misuse taxonomy. We found \newMisuseCases{} new misuse cases from literature and removed two redundant/ambiguous misuse case from the taxonomy,
    thus increasing the number of misuse from $105$ to \totalMisuseCases{} ({\bf RC$_1$}).
    \item {\textbf{Additional Mutation Operators}:} Bolstered by our experience of evaluating the \detectors{} from the previous study, we have extended several mutation-operators to facilitate better evaluation of \detectors{} with the goal of finding flaws. Furthermore, we have created additional mutation operators, which we used in the extended evaluation of \detectors{} ({\bf RC$_2$}).
    \item {\textbf{Extended Evaluation of Additional \detectors{}}:}
    In this extension, we evaluated five \detectors{} from industry, namely, \sonarqube, \snyk, \codiga, \deepsource, and \codeguru). Furthermore, we evaluated the updated versions of all the previously evaluated \detectors with the original and additional mutations, except \xanitizer and \coverity. Moreover, we expanded the base applications for mutation by adding \newMutantApplications{} open source, visible (at least 200 stars in GitHub) applications from the wild. With the combination of new base applications and new mutation operators, we created \newMutations{} new mutants, totaling \newTotalMutations{} mutants, which we used to evaluate both the new version of previously used \detectors{}, and newly acquired \detectors{} in this study.
\end{itemize}
Additionally, we have made several  maintainability and extensibility improvements in \tool{} framework over the original implementation of the \oakland{} paper, which we detail in Section~\ref{sec:implementation}.
\myparagraph{Artifact Release}
To foster further research in the evaluation and development of effective cryptographic misuse detection techniques, and in turn, more secure software, we have released all code and data associated with this paper~\cite{masc-online}.

\vspace{-0.5em}
\section{Motivation and Background}\label{sec:motivation}
\vspace{-0.5em}
Insecure use of cryptographic APIs is the second most common cause of software vulnerabilities after data leaks~\cite{veracode-report}.
To preempt vulnerabilities before software release, non-experts such as software developers or quality assurance teams are likely to use crypto-API misuse detectors (or {\em crypto-detectors}) as a part of the Continuous Integration/Continuous Delivery (CI/CD) pipeline (\eg Xanitizer~\cite{XanitizerRIGSIT} and \shiftleft{}~\cite{shiftleft} used in GitHub Code Scan~\cite{github_third_party_code_scanners}), quality assurance suites (\eg SWAMP~\cite{swamp}) or IDEs (\eg \cognicrypt~\cite{KSA+18}).
Thus, the \textit{inability of a \detector to flag an instance of a misuse that it claims to detect directly impacts the security of end-user software}. %
We illustrate this problem with a motivating example, followed by a threat model that describes the adversarial conditions a \detector may face in the wild.

\subsection{Motivating Example}
\label{sec:motivating-example}
Consider Alice, a Java developer who uses \cryptoguard~\cite{rxa+19}, a state-of-the-art \detector, for identifying cryptographic vulnerabilities in her software before release.
In one of her apps, Alice decides to use the \DES cipher, as follows:
\begin{lstlisting}[frame=tb,caption={\small Instantiating \texttt{DES} as a cipher instance in lower case.}, label={lst:des-misuse-2},language=java]
String a="des"; Cipher c=Cipher.getInstance(a); c.init(Cipher.ENCRYPT_MODE,k,iv);
byte enc = c.DoFinal(text);
\end{lstlisting}
This is another instance of the misuse previously shown in Listing~\ref{lst:des-example}, \ie using the vulnerable \DES cipher. %
{\em \cryptoguard} is unable to detect this vulnerability as Alice uses "{\tt des}" instead of "{\texttt{DES}}" as the parameter (see  Section~\ref{sec:results}).
However, this is a problem, because the lowercase parameter makes no functional difference as Java officially supports both parameter choices.
As \cryptoguard does not detect this vulnerability, Alice will assume that her app is secure and release it to end-users.
Thus, we need to systematically identify such flaws, which would allow the  maintainers of \detectors such as \cryptoguard to promptly fix them, enabling holistic security improvements.

\subsection{Threat Model}\label{sec:threat-model}

To evaluate \detectors, we first define the {\em scope} of our evaluation, for which we leverage the documentation of popular \detectors to understand how they position their tools, \ie what use cases they target (see~\cite{mascArtifact} for all quotes).
For example, \coverity's documentation states that it may be used to \emphdblquote{ensure \underline{compliance} with \underline{security} and coding standards}
.
Similarly, \spotbug's \textit{Find Security Bugs} plugin is expected to be used for \emphdblquote{\underline{security audits}}~\cite{spotbugs_findsecbugs_description}.
Further, \cognicrypt states that its analyses \emphdblquote{\underline{ensure} that \underline{all usages} of cryptographic APIs \underline{remain secure}}~\cite{knr+17}, which may suggest the ability to detect vulnerabilities in code not produced by the developer, but originating in a third-party source (\eg external library, or a contractor), whose developer may not be entirely ``virtuous''.
In fact, 8/9 \detectors evaluated in the \oakland paper, in addition to all the \detectors evaluated in this extended study claim similar cases that demand strong guarantees, \ie for tasks such as compliance auditing or security assurance that are generally expected to be performed by an {\em independent} third party that assumes the worst, including bad coding practices or malpractice~\cite{and20}.
As aptly stated by Anderson~\cite{and20}, \emphdblquote{When you really want a protection property to hold, it's vital that the design and implementation be subjected to hostile review}.

Thus, given that \detectors claim to be useful for \textit{hostile-environment} related tasks, such as compliance audits, it is  likely for them to be deployed in adversarial circumstances, \ie where there is tension between the party that uses a \detector for evaluating software for secure crypto-API use (\eg markets such as Google Play, compliance certifiers such as Underwriters Laboratories (UL)~\cite{iotSecurityRating}), and the party implementing the software (\eg a third-party developer).
With this context, we define a threat model consisting of three types of adversaries (\ref{adversary:benign_accident} -- \ref{adversary:evasive}), which guides/scopes our evaluation according to the conditions \detectors are likely to face in the wild based on their use-case related claims:

\begin{enumerate}[label=\textbf{\bf T$\arabic*$}\xspace,ref=\textbf{T$\arabic*$}\xspace]
\item \label{adversary:benign_accident} {\bf \em Benign developer, accidental misuse -- } This scenario assumes a benign developer, such as Alice, who accidentally misuses crypto-API, but attempts to detect and address such vulnerabilities using a \detector before releasing the software.

\item \label{adversary:benign_fixing} {\bf \em Benign developer, harmful fix --} This scenario also assumes a benign developer such as Alice who is trying to address a vulnerability identified by a \detector in good faith, but ends up introducing a new vulnerability instead.
For instance, a developer may not fully understand the problem identified by a \detector, such as missing certificate verification (\eg an empty \checkServerTrusted method in a custom \trustManager), and address it with an inadequate Stack Overflow fix~\cite{realMisuseStackCondition}.

\item \label{adversary:evasive} {\bf \em Evasive developer, harmful fix -- } This scenario assumes a developer whose goal is to finish a task as quickly or with low effort (\eg a third-party contractor), and is hence attempting to \textit{purposefully} evade a \detector.
Upon receiving a vulnerability alert from a \detector, such a developer may try quick-fixes that do not address the problem, and simply hide it (\eg hiding the vulnerable code in a class that the \detector does not analyze).

For example, Google Play evaluates apps by third-party developers to ensure compliance with its crypto-use policies, but there is ample evidence of developers seeking to actively violate these policies~\cite{bypassGoogleSSLOne, bypassGoogleSSLTwo}.
This adversarial scenario is motivated by \detectors{} that make \textit{hostile}-review related (marketing) \textit{strong} claims, such as compliance and security audits.

In fact, as Oltrogge et al.~\cite{oha+21} recently discovered that developers have been using Android's Network Security Configurations (NSCs) to {\em circumvent safe defaults} (\eg to permit cleartext traffic that is disabled by default).
Note that we do not consider a developer to abuse known limitations, such as documented inclusion of APIs. We elaborate this further in Section~\ref{sec:result-identifying-flaws}.
\end{enumerate}
This threat model, which guides \tools design (Section~\ref{sec:goals}), represents that adversarial conditions under which \detectors may have to operate in practice, and hence, motivates an evaluation based on what \detectors~{\em should be} detecting.
Furthermore, our threat model only considers misuse, including evasive (\ref{adversary:evasive}), that are within the scope of static analysis, \ie misuse that can be detected using techniques such as pattern matching, sensitivities (\eg taint, flow, context, object, field), and/or inter-procedural analysis that can be resolved statically. Techniques outside the scope of static analysis, \eg reflection, foreign function interface, and values that can only be determined at runtime, are considered out of scope in our threat model. Additionally, vulnerable, custom implementations of cryptographic algorithms, such as DES-ECB for Java, are considered outside the scope of our threat model.
However, we note that there may be a gap between {\em what should be} and {\em what is}, \ie while \detectors may want to be relevant in strong deployment scenarios such as compliance checking, their actual design may not account for adversarial use cases (\ie\ \ref{adversary:evasive}).
Therefore, we balance our evaluation that uses this threat model with a discussion that acknowledges all views related to this argument, and especially the tool designer's perspective (Sec.~\ref{sec:discussion}).

\section{Related Work}
\label{sec:related_works}
Security researchers have recently shown significant interest in the external validation of static analysis tools
 ~\cite{QWR18,droidbench,iccbench,PBW18,bkm+18,ZLL+24,BLK24}. %
Particularly, there is a growing realization that static analysis security tools are sound in theory, but {\em soundy in practice}, \ie consisting of a core set of sound decisions, as well as certain strategic unsound choices made for practical reasons such as performance or precision~\cite{lss+15}.
Sound{\em y} tools are desirable for security analysis as their sound core ensures sufficient detection of targeted behavior, while also being {\em practical}, \ie without incurring too many false alarms.
However, given the lack of oversight and evaluation they have faced so far, {\em \detectors} may violate this basic assumption behind sound{\em i}ness and may in fact be {\em unsound}, \ie have fundamental flaws that prevent them from detecting even straightforward instances of crypto-API misuse observed in apps.
This intuition drives our approach for systematically evaluating \detectors, leading to novel contributions that deviate from related work.

To the best of our knowledge, \tool is the first framework to use mutation testing, combined with a large-scale data-driven taxonomy of crypto-API misuse, for comprehensively evaluating the detection ability of \detectors to find design/implementation flaws.
However, in a more general sense, Bonett et al.~\cite{bkm+18} were the first to leverage the intuition behind mutation testing for evaluating Java/Android security tools, and developed the $\mu$SE framework for evaluating data leaks detectors (\eg FlowDroid~\cite{arf+14} and Argus~\cite{wror14}).
\tool significantly deviates from $\mu$SE in terms of its design focus, in order to address the unique challenges imposed by the problem domain of crypto-misuse detection (\ie{} {\bf RC$_1$} -- {\bf RC$_3$} in Sec.~\ref{sec:introduction}).
Particularly, \muse assumes that for finding flaws, it is sufficient to {\em manually} define "a" security operator and strategically place it at hard-to-reach locations in source code.
This assumption does not hold when evaluating \detectors as it is improbable to cast cryptographic misuse as a single mutation, given that cryptographic misuse cases are diverse ({\bf RC$_1$}), and developers may express the same type of misuse in different ways ({\bf RC$_2$}).
For example, consider three well-known types of misuse that would require unique mutation operators: {\sf (1)} using \DES for encryption (operator inserts prohibited parameter names, \eg\ \DES), {\sf (2)} trusting all SSL/TLS certificates (operator creates a malformed \trustManager), and {\sf (3)} using a predictable initialization vector (IV) (operator derives predictable values for the IV).
In fact, developers may even express the same misuse in different ways, necessitating unique operators to express such distinct {\em instances}, \eg the \DES misuse expressed differently in Listing~\ref{lst:des-example} and Listing~\ref{lst:des-misuse-2}.
Thus, instead of adopting \muse's single-operator approach, \tool designs general usage-based mutation operators that can expressively instantiate misuses from our taxonomy of \totalMisuseCases misuses.
In a similar manner, \tools contextualized mutation abstractions (\ie for evaluating \detectors) distinguish it from other systems that perform vulnerability injection for C programs~\cite{dhk+16}, discover API misuse using mutation~\cite{WLW+19,GKL+19}, or evaluate static analysis tools for precision using handcrafted benchmarks or user-defined policies~\cite{QWR18,PBW18}.

Finally, the goal behind \tool is to assist the designers of \detectors~\cite{fhm+12,ebfk13,cognicrypteclipse,knr+17,KSA+18} in identifying design and implementation gaps in their tools, and hence, \tool is complementary to the large body of work in this area.
Particularly, prior work provides rule-sets or benchmarks~\cite{BDA+17,BDA+19} consisting of a limited set of cryptographic ``bad practices''~\cite{BD16}, or taxonomies of smaller subsets (\eg SSL/TLS misuse taxonomy by Vasan et al.~\cite{NVK16}), or examines the precision of \detectors~\cite{CLW+24}.
However, we believe that ours is the first systematically-driven and comprehensive taxonomy of {\em crypto-API} misuse, which captures \totalMisuseCases{} cases that are further expanded upon into numerous unique misuse instances through \tools operators.
Thus, relative to prior handcrafted benchmarks, \tool can thoroughly test detectors with a far more comprehensive set of crypto-misuse instances.

\section{the \tool Framework}
\label{sec:goals}

\begin{figure}[t]
	\centering
    \includegraphics[width=0.8\linewidth]{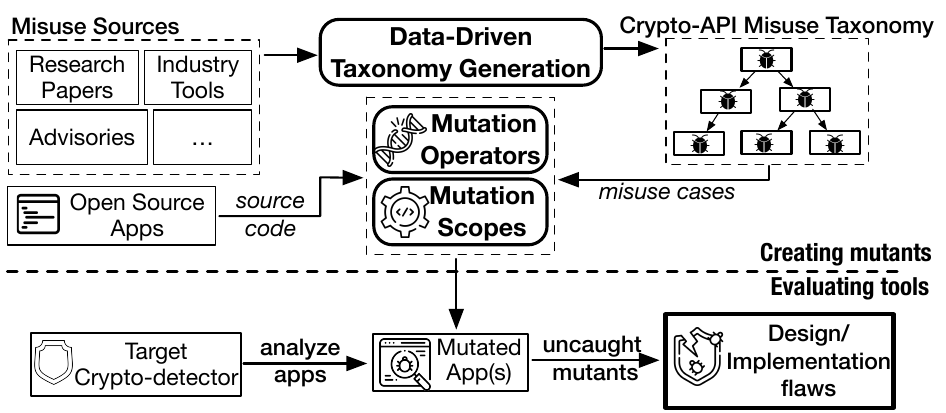}
    \caption{\small A conceptual overview of the \tool framework.}
    \label{fig:tool}
\end{figure}

We propose a framework for Mutation-based Analysis of Static Crypto-misuse detection techniques (or \tool). %
Fig.~\ref{fig:tool} provides an overview of the \tool framework.
As described previously ({\bf RC$_1$}), cryptographic libraries contain a sizable, diverse set of APIs, each with different potential misuse cases, leading to an exponentially large design space.
Therefore, we initialize \tool by developing a {\em data-driven taxonomy of crypto-API misuse}, which grounds our evaluation in a unified collection of misuse cases observed in practice (Sec.~\ref{sec:taxonomy}).

 The misuse cases in the taxonomy must be {\em instantiated} in an {\em expressive manner} to account for the diverse ways for expressing a misuse, \ie\ {\em misuse instances}, that \detectors may face in practice.
 For example, we previously described two ways of encrypting with \DES: (1) providing \DES as a variable in \ciphergetinstance (Listing~\ref{lst:des-example}), or (2) using it in lowercase (Listing~\ref{lst:des-misuse-2}), which both represent something a benign developer might do (\ie threat \ref{adversary:benign_accident}).
To represent all such instances without having to hard-code instantiations for every misuse case, we identify {\em usage-characteristics} of cryptographic APIs (particularly, in JCA), and leverage them to define {\em general}, {\em usage-based mutation operators}, \ie functions that can create misuse instances (\ie\ {\em mutants}) by instantiating one or more misuse cases from the taxonomy (Sec.~\ref{sec:mutation-operators}).

 Upon instantiating mutants by applying our mutation operators to the misuse cases from the taxonomy, \tool\ {\em seeds}, \ie injects, the mutants into real Java/Android applications.
The challenge here is to seed the mutants at specific locations that reflect the threat scenarios described in Sec.~\ref{sec:threat-model}, because \detectors may not only face various instances of misuse cases, but also variations in {\em where} the misuse instances appear, \eg evasive (\ref{adversary:evasive})
developers may attempt to actively hide code to evade analysis.
Thus, we define the abstraction of {\em mutation scopes} that place the instantiated mutants at strategic locations within code, emulating practical threat scenarios  (Sec.~\ref{sec:layers}).
Finally, we analyze these {\em mutated apps} using the target \detector for evaluation (Sec.~\ref{sec:evaluation}), which results in {\em undetected mutants} that can be then inspected to uncover design or implementation flaws (Sec.~\ref{sec:results}).

\section{Taxonomy of Cryptographic Misuse}
\label{sec:taxonomy}

\begin{figure*}[!ht]
	\centering
    \includegraphics[width=0.96\linewidth]{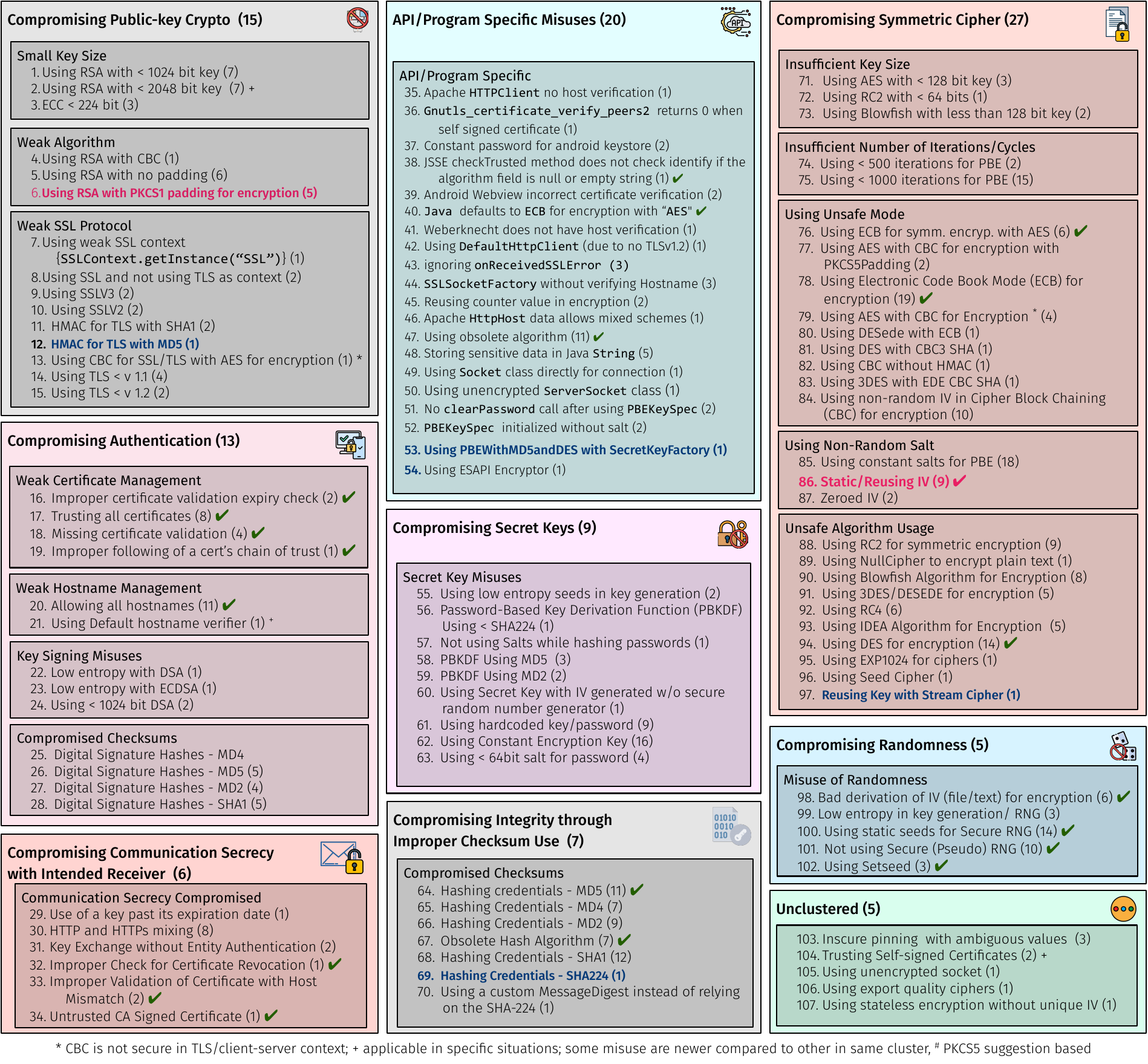}
    \caption{\small The derived taxonomy of cryptographic misuses.  ($n$) indicates misuse was present across $n$ artifacts.  A \checkmark indicates that the specific misuse case was instantiated with \tool's mutation operators for our evaluation (Sec.~\ref{sec:evaluation}). \small{New misuse cases found as part of this extension are highlighted in Blue.} \small Specific misuse IDs {\eg \bf M$_n$} in text refers to $n$th entry, i.e., \misusenumber{1} refers to 1st entry, Using RSA with <1024 bit key.}
    \label{fig:taxonomy}
	\vspace{-1.5em}
\end{figure*}

To ground \tool in real cases of crypto API misuses, we systematically developed a taxonomy that provides a unified perspective of previously-known crypto-API misuse.
Particularly, we focus on identifying instances of crypto-API misuse in the popular and ubiquitous Java ecosystem, as most \detectors that we seek to evaluate were designed to analyze Java/Android apps.

As crypto-API misuse has been widely studied, it is likely that a majority of the misuse cases that we are interested in codifying are already present in existing literature.
Therefore, our methodology identifies crypto-API misuses in existing artifacts sourced from both industry and academia, following Kitchenham et al.'s~\cite{KBB+09} guidelines for identifying relevant artifacts, as well as Petersen et al.'s~\cite{PVK15} recommendations for constructing a systematic mapping, in three main steps:
{\sf (1)} identifying information sources, {\sf (2)} defining the search, inclusion, and exclusion criteria for identifying artifacts, and {\sf (3)} extracting misuse cases from artifacts and clustering them for ease of representation and extensibility.
Two authors executed this methodology and created the taxonomy illustrated in Fig.~\ref{fig:taxonomy}.
Data from each step is provided in the original artifact~\cite{mascArtifact}.

\subsection{Identifying Information Sources}

We considered information sources from both academia and industry.
More specifically, we considered the proceedings of top-tier venues in security and software engineering (\ie USENIX Security, ACM CCS, IEEE S\&P, NDSS, ICSE, ASE, FSE), {\em published after 1999, \ie in the last 20 years}.
Moreover, we also performed a thorough search for relevant keywords (Sec.~\ref{sec:search}) in digital libraries, \ie the ACM Digital Library, IEEE Explore, and Google Scholar, which aided in identifying artifacts that may have fallen outside the top conferences.
Finally, to incorporate sources outside academia, we studied guidelines from the Open Web Application Security Project (OWASP)~\cite{OWASP}, and documentation of several industry tools.

\subsection{Search, Inclusion, and Exclusion Criteria}\label{sec:search}
We select artifacts from the identified sources using a keyword-based search with precise inclusion/exclusion criteria.
We defined 3 classes of keyword phrases and enumerated several keyword combinations for each class, drawing from domain expertise and a random sample of artifacts.

To decide whether to consider an artifact  for further analysis, we defined a simple {\em \textbf{inclusion criterion}}, that the artifact should discuss crypto API misuse or its detection.
We also defined an {\em \textbf{exclusion criterion}}, \ie that the crypto-API misuse described by the artifact relates to a programming environment outside the Java ecosystem, was published prior to 1999, or does not contain relevant information.
Following this methodology, we short-listed 40 artifacts for misuse extraction, \ie\ $35$ from academia and $5$ from industry. Note that we count multiple documents for a single industry tool as one artifact.%

\subsection{Misuse Extraction and Clustering}\label{subsec:misuse_extraction}
Two authors independently extracted, described, and grouped individual misuse cases from the 40 artifacts.
More specifically, each identified misuse case was labeled using a specifically designed data extraction form (see online appendix~\cite{masc-online} for Figure).
Note that since our search included crypto-API misuse from over 20 years, specific API-misuse were introduced over time as part of \textit{cryptographic-hardening}, \eg{} \textit{using RSA with < 1024 bit key (\misusenumber{1})} was disallowed earlier (2013), whereas \textit{using RSA with < 2048 bit key (\misusenumber{2})} became disallowed in 2019. We include all such instances separately in the taxonomy as they are found in the sources. We consider adding all such misuse as found necessary to reflect the state of divergence. For example, some sources explicitly mention using AES with ECB as a misuse (\misusenumber{76}).
In contrast, some sources mention only using ECB as an unsafe mode of operation (\misusenumber{78}), even though historically, ECB is used as a placeholder with the RSA in public-key cryptography for Java.
Similarly, several misuse cases came up relatively fewer number of times, such as \misusenumber{3} was found only thrice in the information sources. Hence, we include the number of times we found a particular misuse in parenthesis beside each entry in the taxonomy, which readers may consider when going through the taxonomy.
Following this process, the two authors met and resolved disagreements, to eventually identify {\em \totalMisuseCases unique misuse cases}.

Such a large set of misuse cases could prove intractable for direct analysis or extension.
Hence, we constructed a {\em categorized} taxonomy by grouping the discovered misuse cases into {\em semantically meaningful clusters}.
Each author constructed the clusters as per two differentiating criteria: {\sf (1)} the {\em security goal/property} represented by the misuse cases (\eg secrecy, integrity, non-repudiation), and {\sf (2)}  its {\em level of abstraction} (\ie specific context) within the communication/computing stack (\eg confidentiality in general, or confidentiality with respect to SSL/TLS).
The two authors met and reached agreement on a taxonomy consisting of {\em 105 misuse cases} grouped into {\em nine semantic clusters}, as shown in Fig.~\ref{fig:taxonomy} in the \oakland{} paper.
The process of taxonomy generation took over {\em two person-months} effort.

\subsection{Extending the Taxonomy}
To expand the taxonomy of crypto-API misuse cases for the current iteration of our work, we applied the existing methodology, covering information sources from both academia and industry, specifically focusing within the year range 2019 - 2022.
Additionally, we manually went through the Common Weakness Enumeration (CWE) database to identify crypto-API misuse cases.
As a result, we were able to identify \newMisuseCases{} crypto-API misuse cases from \newMisuseSources{} additional information sources, as highlighted in blue color in the Taxonomy (Figure~\ref{fig:taxonomy}).

\section{Usage-based Mutation Operators}
\label{sec:mutation-operators}
In designing our mutation operators, we must balance the dichotomous tradeoff between representing as many misuse cases (and their corresponding variations) as possible, while creating a tractable number of operators that can be reasonably maintained in the future.
Thus, building a large set of hard-coded operators that are tightly coupled with specific misuse cases would be infeasible from an engineering/maintenance perspective.
Further, to discover new issues in \detectors, these operators should not exploit general  soundiness-related~\cite{soundinessWebsite,lss+15}) limitations, such as dynamic code execution and implicit calls.
Therefore, we seek to build operators that are general enough to be maintainable, provide expressive instantiation of several misuse cases guided by the threat model in Section~\ref{sec:threat-model}, and without focusing on specific static analysis technique or soundiness issues.

We define the abstraction of {\em usage-based mutation operators}, inspired by a key observation: misuse cases that are unrelated in terms of the security problem %
may still be related in terms of {\em \underline{how} the crypto APIs corresponding to the misuse cases are expected to be used}.
Thus, characterizing the common usage of crypto APIs would allow us to mutate that characterization and define operators that apply to multiple misuse cases, while remaining independent of the specifics of each misuse.

\myparagraph{Common Crypto-API Usage Characteristics}
We identified two common patterns in crypto-API usage by examining crypto-API documentation from JCA, and our taxonomy,
 which we term as {\sf (1)} {\em restrictive} and {\sf (2)} {\em flexible} invocation.
To elaborate, a developer can only instantiate certain objects by providing values from a predefined set, hence the name {\em restrictive} invocation; \eg{} for symmetric encryption with the \ciphergetinstance{} method, JCA {\em only} accepts predefined configuration values for the algorithm name, mode, and padding, in String form.
Conversely, JCA allows significant extensibility for certain crypto APIs, which we term as {\em flexible} invocation; \eg{} developers can customize the hostname verification component of the SSL/TLS handshake by creating a class extending the \hostnameVerifier{}, and overriding its \texttt{verify} method, with {\em any content}.
We leverage these notions of {\em restrictive} \& {\em flexible} usage to define our operator types as follows, with examples in the Appendix~\ref{app:code_snippets}, referred as Listings. We encourage the readers to check Table~\ref{tbl:flaws} and Table~\ref{tbl:flaws_ext} to see concrete use cases of these operators in Section~\ref{sec:results}.

\subsection{Operators based on Restrictive Crypto API Invocation}
\label{sec:restrictive-ops}

Our derived taxonomy indicates that several parameter values used in restrictive API invocations may not be secure (\eg{} \DES, or \MDFIVE).
Therefore, we designed mutation operators ({\bf OPs}) that apply a diverse array of transformations to such values and generate API invocations that are acceptable/compilable as per JCA syntax and conventions, but not secure.

\operator{1}{Atypical case}  This operator changes the case of algorithm specification misuse to an atypical form (\eg{} lowercase), and represents accidental misuse/typos by developers, \eg{} Listing~\ref{lst:des-misuse-2}.

\operator{2}{Weak Algorithm in Variable} This operator represents the relatively common usage of {\em expressing API arguments in variables} before passing them to an API,
  and can be applied to instantiate all misuse cases that are represented by restrictive API invocations, \eg{}
  Listing~\ref{lst:des-example}.

\operator{3}{Explicit case fix}
This operator instantiates misuse using the atypical case for an argument (\eg{} algorithm name) in a restrictive API, as seen in {\bf OP$_1$}, while explicitly {\em fixing} the case.

\operator{4}{Removing noise} This operator extends \opnumber{3} by defining more complex transformations than simple case changes, such as removing extra characters or "noise" present in the arguments.

\operator{5}{Method chains} This operator performs arbitrary transformations (\eg{} among those in {\bf OP$_1$} -- {\bf OP$_4$})  on the restricted argument string, but splits the transformation into a chain of procedure calls, thereby hiding the eventual value (Listing~\ref{lst:method_chain}).

\begin{lstlisting}[frame=tb,caption={{\small Method Chaining (\opnumber{5}).}}, label={lst:method_chain},language=java]
  Class T { String algo="AES/CBC/PKCS5Padding";
  T mthd1(){ algo = "AES";  return this;} T mthd2(){ algo="DES"; return this;} }
  Cipher.getInstance(new T().mthd1().mthd2());
  \end{lstlisting}

\operator{6}{Predictable/Non-Random Derivation}
This operator emulates
deriving a random value
(\ie instead of using a cryptographically-secure random number generator), by performing seemingly complex operations resulting in a predictable value and then using the derived value to obtain other cryptographic parameters, such as IVs (Listing~\ref{lst:bad_derivation_operator}).

We relied on the \ref{adversary:evasive} threat model while creating additional mutation operators in this extension for two reasons.
First, our \oakland{} paper demonstrated that the crypto-API misuse cases \tool{} creates is similar to misuse cases found in the wild. However, the real-life misuse cases can be based on more complex abstractions while still being statically analyzable~\cite{mao+24}.
Second, practitioners expect SASTs, including \detectors{}, to report any vulnerabilities that fall within the scope of program analysis techniques~\cite{ami-fn-oakland24}.
Considering these, we extended several of the existing mutation operators of \tool{} and created additional mutation operators to further improve the evaluation capabilities of \tool{}.
Next, we discuss the mutation-operators built for this extension.

\operator{13}{Iterative Method Chaining} Extending \opnumber{5}, this operator takes a user-specified, arbitrary number ($n$) and creates $n$ methods to be used in chains.
to analyze the internal limits of \detectors{} when following a method chain (Listing~\ref{lst:iterativechaining}).

\operator{14}{Iterative Nested Conditionals}
Extending \opnumber{13}, this operator creates nested conditionals based on user-specified, arbitrary number.
The resulting misuse consists of two nested branches. The "\textit{true}" branch always returns an insecure parameter after traversing, whereas the other returns a secure parameter (Listing~\ref{lst:iterativeconditionals}).

\operator{15}{Method Builder}
  This operator breaks down a crypto-API String parameter into a chain of method calls. It then creates a concatenation method that invokes the character containing methods to chain together the string parameter (Listing~\ref{lst:methodbuilder}).

\operator{16}{Object Sensitivity}
  To evaluate object sensitivity of a \detector{}, this operator creates two objects, \inlinesmall{varA} and \inlinesmall{varB}, with secure and not-secure values in variables respectively, after which the latter is assigned to the former (Listing~\ref{lst:ObjectSensitivity}).

\operator{17}{Build Variable}
This operator breaks a pre-specified value to a character array and passes it to a target crypto API while using the \inlinesmall{toString()} method (Listing~\ref{lst:buildvariable}).

\operator{18}{Substring}
This operator creates a random string containing vulnerable value, \eg{} \inlinesmall{''HelloWorldDES''}.
Next, it uses the \inlinesmall{substring} method to extract the insecure parameter from the original string and passes it to the target crypto-API (Listing~\ref{lst:substring}).

\operator{19}{Constant Value Derivation}
Similar to \opnumber{6}, it derives a constant/predictable value, without using any complex operations (Listing \ref{lst:constantiv}).

\subsection{Operators based on Flexible Crypto API Invocations}
\label{sec:flexible-ops}
In contrast with restrictive APIs, Java allows several types of flexible extensions to crypto APIs represented by interfaces or abstract classes, {\em only enforcing typical syntactic rules}, with little control over what semantics developers express.
Thus, we consider three particular types of flexible extensions developers may make and our {\bf OPs} may emulate, in the context of API misuse cases that involve flexible invocations: {\sf (1)} {\em method overriding}, {\sf (2)} {\em class extension}, and {\sf (3)} {\em object instantiation}.

\subsubsection{Method Overriding}
Crypto APIs are often declared as interfaces or abstract classes containing abstract methods, to facilitate customization.
These abstract classes provide a fertile ground for defining mutation operators with a propensity for circumventing detectors (\ie considering threats~\ref{adversary:evasive}
).

\operator{7}{Ineffective Exceptions} If the correct behavior of a method is to throw an exception, such as invalid certificate, this operator creates misuse instances of two types: {\sf (1)} not throwing any exception, and {\sf (2)} throwing an exception within a conditional block that is only executed when a highly unlikely condition is true,
\eg{} Listing~\ref{lst:conditional_exception}.

\operator{8}{Ineffective Return Values} If the correct behavior of a method is to return a specific value to denote security failure, this operator modifies the return value to create two misuse instances: {\sf (1)} if the return type is boolean, return a predefined value that weakens the crypto, or return {\em null}, and {\sf (2)} introduce a condition block that will always, prematurely return a misuse-inducing value before the secure return statement is reached.
Contrary to \opnumber{7}, this operator ensures that the condition will always return the value resulting in misuse (Listing~\ref{lst:condition_return}).

\operator{9}{Irrelevant Loop} This operator adds loops that {\em seem to perform} security checks before returning a value, but in reality {\em do nothing to change the outcome}
(Listing~\ref{lst:irrelevant_loop_trustmanager}).
\subsubsection{Class Extension}
Creating abstractions on top of previously-defined crypto classes is fairly common; \eg{} the abstract class \inlinesmall{X509ExtendedTrustManager} implements the interface \xtrustManager in the JCA, and developers are expected to extend/implement both these to customize certificate verification.
Similarly, developers may create abstract subtypes of an interface or abstract class; \eg{} Listing~\ref{lst:abstract_extension_empty}
(Appendix~\ref{app:code_snippets}),
Our next set of operators is motivated by this observation:

\operator{10}{Abstract Extension {\em with} Override} This operator creates an abstract sub-type of the parent class (\eg{} Listing~\ref{lst:abstract_extension_empty}), but this time, overrides the methods {\em incorrectly}, \ie adapting the techniques in \opnumber{7} -- \opnumber{9} for creating various instances of misuse.

\operator{11}{Concrete Extension {\em with} Override} This  operator creates a concrete class based on a crypto API, {\em incorrectly} overriding methods similar to \opnumber{10}.

\subsubsection{Object Instantiation Operators}
In Java, objects can be created by calling either the default or a parametrized constructor, while possibly overriding the properties of the object through Inner class object declarations.
We leverage these properties to create \opnumber{12} as follows:

\operator{12}{Anonymous Inner Object}
Creating an instance of a flexible crypto API through constructor or anonymous inner class object is fairly common, as seen for \hostnameVerifier in Oracle Reference Guide~\cite{jcaHostnameVerifierExample} and Android developer documentation~\cite{androidHostnameVerifierExample}, respectively.
Similarly, this operator creates anonymous inner class objects from abstract crypto APIs, and instantiates misuse cases by overriding the abstract methods using \opnumber{7} -- \opnumber{9} (Listing~\ref{lst:abstract_inner_class}).

\tools{} \totaloperators{}  operators are capable of instantiating \textbf{68/\totalMisuseCases} ($\approx 64\%$), misuse cases distributed across \textit{all} 9 \textit{semantic clusters}.
This indicates that \tools operators can express a \textit{diverse majority} of misuse cases, signaling a reasonable trade-off between the number of operators and their expressivity.
Of the remaining 39 cases that our operators do not instantiate, 16 are trivial to implement \eg{} using AES with $<$ 128 bit key in Listing~\ref{lst:trivial}.
Finally, 22 cases ($\approx 21\%$) would require a non-trivial amount of additional engineering effort; \eg{} designing an operator that uses a custom \messageDigest algorithm instead of a known standard, such as \texttt{SHA-3},  would require implementing a custom hashing algorithm.

\section{Threat-based Mutation Scopes}
\label{sec:layers}

We seek to emulate the typical placement of vulnerable code by benign (\ref{adversary:benign_accident},~\ref{adversary:benign_fixing}), and evasive (\ref{adversary:evasive})
developers, for which we
design three {\em mutation scopes}:

\myparagraph{1. {\em Main} Scope}
The \textit{main scope} is the simplest of the three, and seeds mutants at the beginning of the main method of a simple Java or Android template app developed by the authors (\ie instead of the real, third-party applications mutated by the other two scopes).
This specific seeding strategy ensures that the mutant would always be reachable and executable, emulating basic placement by a benign developer (\ref{adversary:benign_accident},~\ref{adversary:benign_fixing}).

\myparagraph{2. {\em Similarity} Scope}
The {\em similarity scope} seeds security operators at locations in a target application's source code where a similar API is already being used, \ie akin to modifying existing API usages and making them vulnerable.
Hence, this scope emulates code placement by a typical, well-intentioned, developer (\ref{adversary:benign_accident},~\ref{adversary:benign_fixing}), assuming that the target app we mutate is also written by a benign developer.
This helps us evaluate if the \detector is able to detect misuse at {\em realistic} locations.

\myparagraph{3. {\em Exhaustive} Scope}
As the name suggests, this scope {\em exhaustively} seeds mutants at {\em all locations} in the target app's code, \ie class definitions, conditional segments, method bodies and anonymous inner class object declarations.
Note that some mutants may not be seeded in all of these locations; \eg a \ciphergetinstance{} is generally surrounded by \trycatch block, which cannot be placed at class scope.
This scope emulates placement by an evasive developer (\ref{adversary:evasive}).

\section{Implementation}
\label{sec:implementation}

The first iteration of \tool, as discussed in our \oakland{} paper, involved three components, namely, {\sf ($1$)} selecting misuse cases from the taxonomy for mutation, {\sf ($2$)} implementing mutation operators that instantiate the misuse cases, and {\sf ($3$)} seeding/inserting the instantiated mutants in Java/Android source code at targeted locations.

\myparagraph{1. Selecting misuse cases from the Taxonomy}
We chose \implementedMisuseCases misuse cases from the taxonomy for mutation with \tools\ \totaloperators operators (indicated by a $\checkmark$ in Fig.~\ref{fig:taxonomy}), focusing on ensuring broad {\em coverage} across our taxonomy as well as on their {\em prevalence}, \ie prioritizing misuse cases that are discussed more frequently in the artifacts used to construct the taxonomy, and which most \detectors can be expected to target,
while being
influenced by the key observation in Section 6, \ie{} APIs that are unrelated may still be related in how they are invoked.
To elaborate,
consider the misuse instances of \ciphergetinstance and \messageDigestInstance. As string-based restrictive APIs,
they are representatives or components of several additional clusters,
such as
Weak SSL protocol (\eg{} \misusenumber{7}-\misusenumber{15}, instantiated using
    \inlinesmall{Mac.getInstance(<parameter>)},
    and \inlinesmall{SSLContext.getInstance(<parameter>)}),
Weak Algorithm (\eg{} \misusenumber{4}-\misusenumber{6} using
\ciphergetinstance)
        from Compromising Public-key Crypto cluster,
Compromised Checksums (\eg{} \misusenumber{25}-\misusenumber{28} using
    \messageDigestInstance{})
        from Compromising Authentication cluster,
Compromising Secret Keys (\eg{} \misusenumber{56},
    \misusenumber{58}, \misusenumber{59} using
    \inlinesmall{Factory.getInstance(<parameter>)}),
Compromising Integrity through Improper Checksum Use cluster
    (\eg{} \misusenumber{64}-\misusenumber{69}) using
        \messageDigestInstance{}, and
Unsafe algorithm usage (\eg{} \misusenumber{88}-\misusenumber{96} using
    \inlinesmall{Cipher.getInstance(<parameter>)}) from Compromising Symmetric Cipher cluster.
If a \detector{} can not detect an identically mutated set of misuse instances
based on both \ciphergetinstance{} and \messageDigestInstance{},
we infer this due to a design flaw.
That is, as the target detector applies the same/similar detection techniques,
it would be unable to detect mutants that are based on
similar, restrictive APIs,
such as \inlinesmall{Mac.getInstance(<parameter>)},
\inlinesmall{SecretKeyFactory.getInstance(<parameter>)}, and
\inlinesmall{SSLContext.getInstance(<parameter>)}.
Otherwise, it is likely that the disparate detection of similarly mutated
misuse instances is due to an implementation flaw.
This approach lets us detect both
flaws tied to specific APIs and
generic patterns of flaws
that are not necessarily tied to specific APIs,
which we later elaborate on in Section~\ref{sec:results}.

\myparagraph{2. Implementing mutants} The mutation operators described in Sec.~\ref{sec:mutation-operators} are designed to be applied to one or more crypto APIs, for instantiating specific misuse cases.
To ensure compilable mutants by design, \tool carefully considers the syntactic requirements of the API being implemented (\eg{} the requirement of a surrounding try-catch block with appropriate exception handling),
as well as the semantic requirements of a particular misuse being instantiated, such as the need to throw an exception only under a truly improbable condition (\eg{} as expressed in {\bf OP$_7$}).
\tool uses Java Reflection to determine the ``syntactic glue'' for automatically instantiating a compilable mutant, \ie exceptions thrown, requirements of being surrounded by a try-catch block, the need to implement a certain abstract class and certain methods, etc. %
\tool then combines this automatically-derived and compilable syntactic glue with parameters, \ie values to be used in arguments, return statements, or conditions, which we manually define for specific operators (and misuse cases), to create mutants.

To further ensure compile-ability and evaluation using only compilable mutants, we take two steps: {\sf ($1$)} We use Eclipse JDT's AST-based checks for identifying syntactic anomalies in the generated mutated apps, and {\sf ($2$)} compile the mutated app automatically using build/test scripts \textit{provided with the original app}.
In the end, every single mutant analyzed by the target \detector is compilable and accurately expresses the particular misuse case that is instantiated.
This level of automation allows \tool to create thousands of mutants with very little manual effort, and makes \tool extensible to future evolution in Java cryptographic APIs (addressing {\bf RC$_3$}).

\myparagraph{3. Identifying Target Locations and Seeding Mutants}
To identify target locations for the similarity scope, we extended the \texttt{MDroid+}~\cite{mtb+18,lbt+17} mutation analysis framework, by retargeting its procedure for identifying suitable mutant locations, adding support for dependencies that crypto-based mutations may introduce, and enabling identification of anonymous inner class objects as mutant-seeding locations, resulting in $10$ additional, custom AST- and string-based location detectors.
Further, \tool{} extends \muse~\cite{bkm+18} to implement the \textit{exhaustive} scope, \ie to identify locations where crypto-APIs can be feasibly inserted to instantiate compilable mutants (\eg{}a \ciphergetinstance has to be contained in a \trycatch block, making it infeasible to insert at class-level).

Based on our prior experience of developing and using the implementation of the MASC framework, we determined several design goals, namely Diversity of Crypto-APIs~\newdg{1}, Open to Extension~\newdg{2}, Ease of evaluating \detectors~\newdg{3}, and adapting to users with different levels of skills~\newdg{4}, as we demonstrated at the FSE'23~\cite{asr-masc-demo23}.
\begin{figure}[tbp]
	\centering
    \includegraphics[width=.7\textwidth]{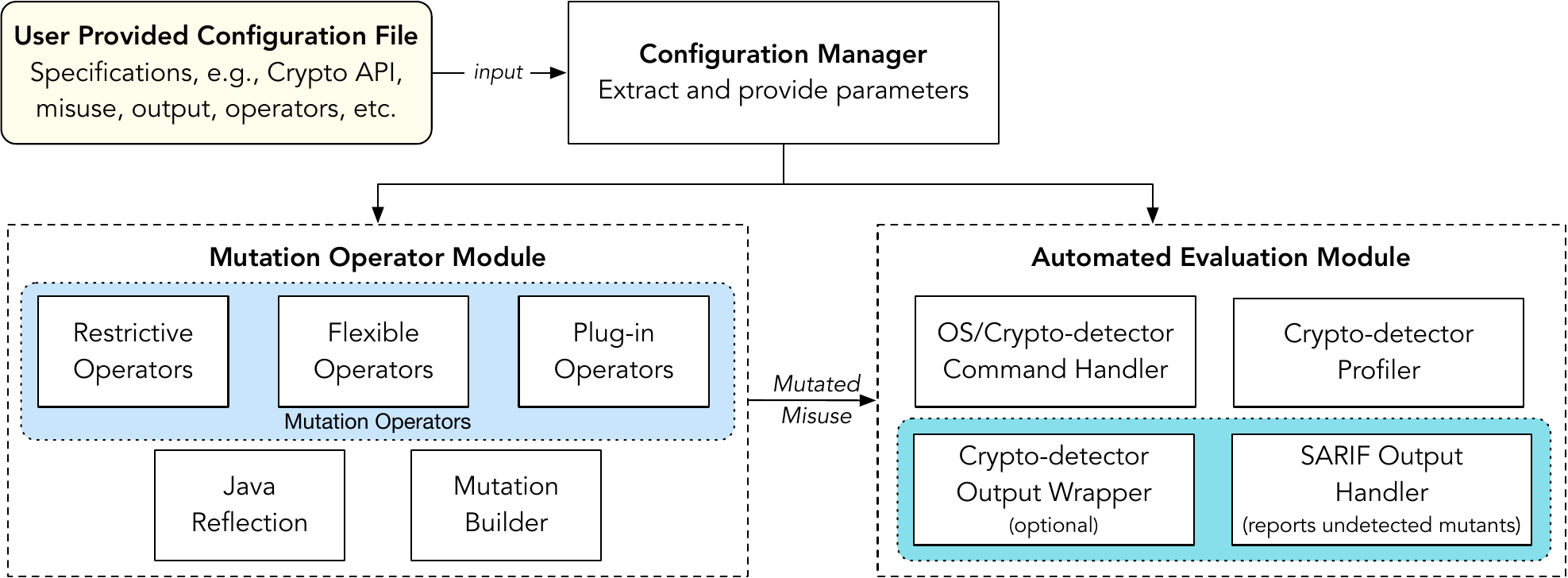}
        \caption{\small Architecture Overview of the Main Scope of \tool{}}
    \label{fig:tool-main-scope}
\end{figure}

To satisfy these design goals (\dgref{1}--\dgref{4}), we implemented \tool{} ($11K+$ effective Java source line of code) following single-responsibility principle across modules, classes, and functions.
While current implementation of \tool{} inherits the \textit{mutation scopes} of the original implementation with internal structural changes, the bulk of the changes with new features in the current implementation of \tool{} are based on the \textit{Main Scope}.
Therefore, we describe the implementation details of \tool{} with a focus on \textit{Main Scope} in the context of the design goals and provide an overview of the architecture in Figure~\ref{fig:tool-main-scope}.

\myparagraphnew{Configuration Manager}
\begin{lstlisting}[basicstyle=\ttfamily\scriptsize,float,numbers=none,caption={{\small
    Example configuration file for \tool{}}},belowcaptionskip=-5mm,label=lst:example_config,emphstyle=\bfseries, language=bash]
scope = main # name of the scope to be used for mutation
type = StringOperator #name of the operator to be used for mutation
outputDir = app/outputs
apiName = javax.crypto.Cipher
invocation = getInstance # Method call from crypto-API
secureParam = AES/GCM/NoPadding # Secure parameter to use with crypto-API
insecureParam = AES # insecure parameter to use with crypto-API
noise = ~ # noise value used with mutation
variableName = cryptoVariable # variable, class name used to create necessary structures
className = CryptoTest
appName = <Name of the App> # name of the app for similarity-scope specific mutation
\end{lstlisting}

To make \tool{} as flexible as possible, we decoupled the crypto-API specific parameters from the internal structure of \tool{}.
As a result, user can specify any crypto-API along with its necessary parameters through an external configuration file defining the base crypto-API misuse case. The configuration file follows a key-value format, as shown in Listing~\ref{lst:example_config}.
Additionally, user can specify the mutation operators and scope to be used, along with other configuration values, thus satisfying \dgref{1}.

\myparagraphnew{Mutation Operator Module}
\tool{} analyzes the specified crypto-API and uses the values specified by the user (\eg{}secure, and insecure parameters to be used with the API) for creating mutated crypto-API misuse instances.
Internally, the decoupling of crypto-APIs from \tool{} is made possible through the use of \textit{Java Reflection} based API analysis and Java Source Generation using the \textit{Java Poetry} Library (\dgref{1}).
While both the original and current implementation of \tool{} comes with several generalizable mutation operators, the current implementation of \tool{} includes an additional plug-in structure that facilitates creating custom mutation operators and custom key-value pairs for the configuration file.
Both of these can be done \textit{externally}, \ie no modification to source code of \tool{} is necessary (\dgref{2}). We provide additional details about \tools{} mutation operators in Section~\ref{sec:mutation-operators}.

\myparagraphnew{Automated Evaluation Module}
The current implementation of \tool{} leverages the SARIF formatted output to automate evaluation of \detectors{}.
To make end-to-end analysis automated, \tools{} can be configured to use \detector{} specific commands, such as \eg{}compiling a mutated source code for analysis, evaluation stop conditions, command for running \detector{}, output directory, and more (\dgref{3}--\dgref{4}).

Finally, \tool{} is implemented to produce verbose logs.
With the combination of flexible configuration, it is therefore possible to use the stand-alone binary \tool{} jar file as a module of another software. As a proof of concept, we implemented \mascweb{}, a \textit{python-django} based front-end that offers all the functionalities of the \tool{}
that uses the binary jar of \tool{} as a module (\dgref{4}).

\section{Evaluation Overview and Methodology}
\label{sec:evaluation}

The two main goals of our evaluation are to (1) measure the effectiveness of \tool at uncovering flaws in \detectors, and (2) learn the characteristics of the flaws and their real-world impact, in terms of the  security of end-user applications.
Therefore, we formulate the following research questions:

\begin{enumerate}[label=$\bullet$ \textbf{RQ$_\arabic*$:},ref=\textbf{RQ$_\arabic*$},wide, labelindent=0pt]
\item \label{rq:canfind}{\textit{Can \tool{} discover flaws in \detectors?}}
\item \label{rq:classes} {\textit{What are the characteristics of these flaws?}}
\item \label{rq:impact}{\textit{What is the impact of the flaws on the effectiveness of \detectors in practice?}}
\end{enumerate}

To answer \ref{rq:canfind} -- \ref{rq:impact}, we performed experiments throughout several years, spanning from 2021 to 2024.
As part of our \oakland{} paper, we first used \tool to evaluate a set of {\em nine} major \detectors, namely \cryptoguard, \cognicrypt, \xanitizer, \coverity, \spotbugfull, \qark, \codeqllgtm, \codeqlgcs (GCS), and \shiftleft Scan, prioritizing practically relevant tools that were recent and under active maintenance.
As part of this study, we evaluated all of these \detectors{} with their updated versions, except \coverity{} and \xanitizer{},  in addition to five new \detectors, namely \codeguru, \snyk, \deepsource, \codiga, and \sonarqube, totaling $12$ \detectors.

To be consistent, we addressed several changes in \detectors{} life-cycle during this study.
First, \codeqllgtm{} was merged with \codeqlgcs{} and ceased to exist separately. As \codeqlgcs{} still stores the test suites from \codeqllgtm{} suites, we used \codeqlgcs{} with \textit{lgtm} test suite as a proxy for the \codeqllgtm{} service.
Furthermore, \codeqlgcs{} has two regularly maintained, security-specific test suites~\footnote{https://docs.github.com/en/code-security/code-scanning/managing-your-code-scanning-configuration/codeql-query-suites}.
First, the \textit{default} test suite is described to be more precise by giving "\textit{fewer low confidence}" results to reduce false-positives. Second, \textit{security-extended} is optionally available, and consists of all rules from the \textit{default} test suite and additional rules with lower precision. Since we are interested in identifying flaws in \detectors, we chose \textit{security-extended} test suite for this extended study for evaluating \codeqlgcs{}.
Finally, several of these new \detectors{} are web-based and can only be used as part of CI/CD workflow. As those \detectors{} do not offer a "fixed version" for evaluation, we used the latest version whenever possible, and confirmed the existence of the flaws as of \textit{April 2024}.

As \tools usefulness is in systematically evaluating individual \detectors by characterizing their detection-ability, with the goal of enabling improvement in the tools, and hence, the results of our evaluation indicate gaps in individual tools, and not comparative advantages.

\myparagraphnew{Step 1 -- Selecting and mutating apps} During the first iteration of this study for our \oakland{} paper, we used \tool to mutate \testedAndroidApps open source Android apps from F-Droid~\cite{fdroid} and Github~\cite{Github}, and four sub-systems %
of \qpid~\cite{qpid}, a large Java Apache project (for list see~\cite{mascArtifact}).
Our selection was biased towards popular projects that did not contain obsolete dependencies (\ie  compilable using Android Studio 4/Java LTS 1.8).
Moreover, we specifically used the \similarityScope on 3/\testedAndroidApps Android apps, and all 4 Qpid sub-systems, as they contained several types of crypto API usage (\eg \cipher, \messageDigest, \xtrustManager).
In total, we generated \totalMutationsAndroid mutants using the \testedAndroidApps Android apps (producing \texttt{\small src} and \texttt{\small apk}) and \totalMutationsJava mutants using the \testedJavaComponents Java programs (producing \texttt{\small src} and \texttt{\small jar}), totaling \totalMutations mutants.
We confirmed that each mutated app was compilable and runnable.
Generating these {\em $20$}k mutants took \tool roughly $15$ minutes, and did not require any human intervention, addressing {\bf RC$_3$}.
As the cost to generate this volume of mutants is feasible, \tool may not benefit from generating a focused subset of mutants (as we detail in \oakland
).

In this extension of the study, for mutation using \tool{}, we selected 10 additional Android applications for the exhaustive scope, four Android applications for the selective scope, and one Java-based application from Google.
Similar to the previous study, these apps were chosen based on few specific criteria to represent popular, regularly updated applications, such as having more than 200 stars in GitHub for Android-based projects, and latest commit dating back to 2022, and having no obsolete dependencies.
Additionally, we relied on GitHub's advance code search feature to identify Android applications with existing crypto-API calls to ensure that those apps can be mutated using \tools{} selective scope.
Finally, we manually confirmed that those 5 newly selected applications indeed contained crypto-API call sites.

Using the extended \tool{} with its additional mutation operators, we mutated these 15 applications and generated \newMutations{} new mutants. We used  \totalMutations{} from the previous study, and \newMutations{} from this study, totaling \newTotalMutations{} mutations, for evaluating the \detectors{}.

\myparagraphnew{Step 2 -- Evaluating \detectors and identifying {\em undetected} mutants}
To evaluate a \detector, we analyzed the mutants using the \detector, and identified the mutants that were {\em not detected} by it, \ie\ {\em undetected} as misuse using similar methodology in both the previous and current study.
To facilitate accurate identification of undetected mutants, we compare the mutation log \tool{} generates when inserting mutants (which describes the precise location and type of mutant injected, for each mutant), and the reports from \detectors{}, which for all the tools contained precise location features such as the class/method names, line numbers, and other details such as associated variables.
To elaborate, we use the following methodology:
We first compare the analysis report generated by the \detector/target on a mutated app, with its analysis report on the original (\ie unmutated) version of the same app.
Any difference in the two reports can be attributed to mutants inserted by \tool, \ie denotes an instance of a ``mutant being not detected''.
To identify the mutant that was not detected, we obtain location features (\ie type, file, class, method, line number and/or variables associated) of the specific ``undetected mutant'' from the \detectors report on the mutated app, and use them to search for a unique mutant in \tools mutation log.
If a match is found, we mark that specific mutant as not detected.
We additionally confirm the location of each undetected mutant by referring to the mutated app's source code.
Once all the undetected mutants are identified, the remaining mutants (\ie inserted in \tools mutation but not detected) are flagged as undetected.

This approach ensures that alarms by the \detector for anomalies not inserted by \tool are not considered in the evaluation, and all mutants inserted by \tool are identified as either detected or undetected.
Our semi-automated implementation of this methodology, which adapts to the disparate report formats of the studied \detectors, is detailed in \oakland{}.

\myparagraphnew{Step 3 -- Identifying {\em flaws}~(\ref{rq:canfind})}\label{sec:result-identifying-flaws}
We analyzed over $400$ randomly chosen, undetected mutants to discover {\em flaws}, wherein a flaw is defined as a misuse case that a particular \detector{} {\em claims} to detect in its documentation, but fails to detect in practice.
We took care to also exempt the exceptions/limitations explicitly stated in the \detector's documentation, ensuring that all of our identified flaws are {\em novel}. %
On a similar note, while a crypto-detector may seem flawed because it does not detect a newer, more recently identified misuse pattern, we confirm that all the flaws we report are due to misuse cases that are older than the tools in which we find them.
This can be attributed to two features of our evaluation: our choice of the most frequently discussed misuse cases in a taxonomy going back 20 years and our choice of tools that were recently built or maintained.
Finally, to confirm the flaw without the intricacies of the mutated app, we created a corresponding minimal app that contained only the undetected misuse instance (\ie the flaw), and re-analyzed it with the detector.

\myparagraphnew{Step 4 -- Characterizing flaws (\ref{rq:classes})}
We characterized the flaws by grouping them by their most likely cause, into {\em flaw classes}.
We also tested each of the tools with the minimal examples for all the flaws, allowing us to further understand each flaw given its presence/absence in certain detectors, and their documented capabilities/limitations.
We reported the flaws to the respective tool maintainers, and contributed 3 patches to CryptoGuard that were all integrated~\cite{mascArtifact}.

\myparagraphnew{Step 5 -- Understanding the practical {\em impact} of flaws~(\ref{rq:impact})}
To gauge the impact of the flaws, we studied publicly available applications
\textit{after} investigating \ref{rq:canfind} and \ref{rq:classes} during the previous study.
We first tried to determine if the misuse instances similar to the ones that led to the flaws were also found in real-world apps (\ie public GitHub repositories) using GitHub Code Search~\cite{github-search}, followed by manual confirmation.
Additionally, we manually searched Stack Overflow~\cite{stackoverflow} and Cryptography Stack Exchange~\cite{cryptostackexchange} for keywords such as "unsafe hostnameverifier" and "unsafe x509trustmanager".
Finally, we narrowed our search space to identify the impact on apps {\em that were in fact analyzed with a \detector}.
As only \codeqllgtmshort showcases the repos that it scans, we manually explored the top $11$ Java repositories from this set (prioritized by the number of contributors), and discovered several misuse instances that \codeqllgtmshort tool may have failed to detect due to the flaws discovered by \tool.

\myparagraphnew{Step 6 -- Attributing flaws to mutation vs. base instantiation}
To determine whether a flaw detected by \tool can be attributed to mutation, versus the \detector's inability to handle even a base case (\ie the most literal instantiations of the misuse case from the taxonomy),
we additionally evaluated each \detector with base instantiations for each misuse that led to a flaw exhibited by it.

\section{Results and Findings}\label{sec:results}
Our manual analysis of undetected mutants revealed \countflaws flaws in our previous \oakland{} study, and six additional flaws in this study across \detectors we evaluated, which we resolved to both design and implementation-gaps (\ref{rq:canfind}).

    \newcolumntype{Y}[1]{>{\raggedright\arraybackslash}p{#1}}
    \begin{normalsize}
    \centering
    \begin{longtable}{l Y{1.5in} p{3.6in}}
    \caption{Descriptions of Flaws discovered by Analyzing \detectors{}}
    \label{tbl:flaws}\\
    \toprule
    \textbf{ID} & \textbf{Flaw Name}
    \linebreak (\textbf{\footnotesize Operator, Base Misuse IDs}) &
    \textbf{Description of Flaws} \\
    \endfirsthead
    \toprule
    \textbf{ID} & \textbf{Flaw Name}
    \linebreak (\textbf{\footnotesize Operator, Base Misuse IDs}) &
    \textbf{Description of Flaws}  \\
    \midrule
    \endhead
    \multicolumn{3}{c}{\textit{Continued on next page.}}\\
    \endfoot
    \multicolumn{3}{p{\textwidth}}{+ flaws were observed for multiple API misuse cases,
    * Certain seemingly-unrealistic flaws may be seen in or outside a \detector's ``scope'', depending on the perspective; see Section~\ref{sec:discussion} for a broader treatment of this caveat.}\\
    \endlastfoot
    \midrule
    \multicolumn{3}{p{\textwidth}}{\textsc{\textbf{\fcdifferentcase+}}}\\
\midrule
    \flawtag{F1}{flaw:smallCaseParameter} &
    smallCaseParameter \linebreak  (\opnumber{1}, \misusenumber{40, 47, 76, 78, 94}) &
    Not detecting an insecure algorithm provided in lower case;
    \eg{} \inline{Cipher.getInstance("des");}
    \\
\midrule
    \multicolumn{3}{p{\textwidth}}{\textsc{\textbf{\fcvalueresoluion+}}}\\
\midrule

    \flawtag{F2}{flaw:valueInVariable} &
    value in variable
    \linebreak (\opnumber{2}, \misusenumber{40, 47, 76, 78, 94}) &
    Not resolving values passed through variables. \eg{}
    \inline{String value = "DES"; Cipher.getInstance(value);}
    \\
\midrule

    \flawtag{F3}{flaw:secureParameterReplaceInsecure}* &
    secure parameter replaced by not secure
    (\opnumber{4}, \misusenumber{64, 67}) &
    Not resolving parameter replacement; \eg{}\
    \inline{MessageDigest.getInstance( "SHA-256".replace("SHA-256", "MD5"));}
    \\
\midrule

    \flawtag{F4}{flaw:insecureParameterReplaceInsecure}* &
    not secure parameter replaced by not secure
    \linebreak (\opnumber{4}, \misusenumber{40, 47, 76, 78, 94}) &
    Not resolving an {\em unsafe} parameter's replacement with another {\em unsafe} parameter \eg{} \newline
    \inline{Cipher.getInstance("AES".replace("A", "D"));} (\ie where "AES" by itself is insecure as it defaults to using ECB).
    \\
\midrule

    \flawtag{F5}{flaw:stringCaseTransform}* &
    string case transform
    \linebreak (\opnumber{3}, \misusenumber{40, 47, 76, 78, 94}) &
    Not resolving the case after transformation for analysis; \eg
    \inline{Cipher.getInstance("des".toUpperCase( Locale.English));}
    \\
\midrule

    \flawtag{F6}{flaw:noiseReplace}* & noise replace
    \linebreak (\opnumber{4}, \misusenumber{40, 47, 76, 78, 94}) &
    Not resolving noisy versions of insecure parameters, when noise is removed through a transformation; \eg{} \newline
    \inline{Cipher.getInstance("DE\$S".replace("\$", ""));}
    \\
\midrule

    \flawtag{F7}{flaw:parameterFromMethodChaining} &
    parameter from method chaining
    \linebreak (\opnumber{5}, \misusenumber{40, 47, 76, 78, 94}) &
    Not resolving insecure parameters that are passed through method chaining, \ie from a class that contains both secure and insecure values; \eg{}
    \inline{Cipher.getInstance(obj.A().B().getValue());}
    where obj.A().getValue() returns the secure value, but obj.A().B().getValue(), and obj.B().getValue()  return the insecure value.
    \\
\midrule

    \flawtag{F8}{flaw:deterministicByteFromCharacters}* &
    deterministic byte from characters
    \linebreak (\opnumber{6}, \misusenumber{86}) &
    Not detecting {\em constant IVs}, if created using complex loops, casting, and string transformations;
    \eg{} a
    \inline{new IvParameterSpec(v.getBytes(),0,8)}, which uses a
    \inline{String v=""; for(int i=65; i<75; i++)\{ v+=(char)i;\}}
    \\
\midrule

    \flawtag{F9}{flaw:predictableByteFromSystemAPI} &
    predictable byte from system API
    \linebreak (\opnumber{6}, \misusenumber{86}) &
    Not detecting {\em predictable IVs} that are created using a predictable source (\eg{} system time), converted to bytes;
    \eg
    \inline{new IvParameterSpec(val.getBytes(),0,8);},
    such that
    \inline{val = new Date(System.currentTimeMillis()).toString();}
    \\
\midrule

    \multicolumn{3}{p{\textwidth}}{\textsc{\textbf{\fcomplexinheritance}}}\\
\midrule

    \flawtag{F10}{flaw:X509ExtendedTrustManager} &
    X509ExtendedTrustManager
    \linebreak (\opnumber{12}, \misusenumber{16-19, 38}) &
    Not detecting {\em vulnerable SSL verification} in {\em anonymous inner class objects} created from the {\scriptsize \tt X509ExtendedTrustManager} class from JCA; \eg{} see Listing~\ref{lst:aic_X509ExtendedTrustManager} in Appendix).
    \\
\midrule

    \flawtag{F11}{flaw:X509TrustManagerSubType} &
    X509TrustManager SubType
    \linebreak (\opnumber{12}, \misusenumber{16-19, 38}) &
    Not detecting {\em vulnerable SSL verification} in anonymous inner class objects {\em created from an empty abstract class} which implements the {\scriptsize \tt X509TrustManager} interface; \eg{} see Listing~\ref{lst:aic_x509tm}).
    \\
\midrule

    \flawtag{F12}{flaw:IntHostnameVerifier} &
    Interface of Hostname Verifier
    \linebreak (\opnumber{12}, \misusenumber{20}) &
    Not detecting {\em vulnerable hostname verification} in an anonymous inner class object that is created from an {\em interface that extends} the {\tt \scriptsize HostnameVerifier} interface from JCA; \eg{} see Listing~\ref{lst:aic_empty_ext_interface_hostname} in Appendix.
    \\
    \midrule

    \flawtag{F13}{flaw:AbcHostnameVerifier} &
    Abstracted Hostname Verifier
    \linebreak (\opnumber{12}, \misusenumber{20}) &
    Not detecting {\em vulnerable hostname verification} in an anonymous inner class object that is created from an {\em empty abstract class} that {\em implements} the {\tt \scriptsize HostnameVerifier} interface from JCA; \eg{} see Listing~\ref{lst:aic_empty_ext_abstract_hostname} in Appendix.
    \\
\midrule

    \multicolumn{3}{p{\textwidth}}{\textsc{\textbf{\fcgenericnoise}}}\\
\midrule

    \flawtag{F14}{flaw:X509TrustManagerGenericConditions} &
    X509TrustManager Generic Conditions
    \linebreak (\opnumber{7}, \opnumber{9}, \opnumber{12}, \misusenumber{16-19, 38}) &
    Insecure validation of an overridden {\scriptsize \tt checkServerTrusted} method created within an anonymous inner class (constructed similarly as in {\bf F13}), due to the failure to detect {\em security exceptions thrown under impossible conditions}; \eg{} {\scriptsize \tt  if(!(true||arg0 == null||arg1 == null)){ throw new CertificateException();}}
    \\
\midrule

    \flawtag{F15}{flaw:IntHostnameVerifierGenericCondition} &
    Interface Hostname Verifier Generic Conditions
    \linebreak (\opnumber{8}, \opnumber{12}, \misusenumber{20}) &
    Insecure analysis of vulnerable hostname verification, \ie the {\tt \scriptsize verify()} method within an anonymous inner class (constructed similarly as in {\bf F14}), due to the failure to detect {\em an always-true condition block that returns {\tt \scriptsize true}}; \eg{} {\scriptsize \tt if(true || session == null) return true; return false;}
       \\
\midrule

    \flawtag{F16}{flaw:AbcHostnameVerifierGenericCondition} &
    Abstract Hostname Verifier Generic Conditions
    \linebreak (\opnumber{8}, \opnumber{12}, \misusenumber{20}) &
    Insecure analysis of vulnerable hostname verification, \ie the {\tt \scriptsize verify()} method within an anonymous inner class (constructed similarly as in {\bf F15}), due to the failure to detect {\em an always-true condition block that returns {\tt \scriptsize true}}; \eg{} {\scriptsize \tt if(true || session == null) return true; return false;}
    \\
    \midrule
    \multicolumn{3}{p{\textwidth}}{\textsc{\textbf{\fcspecificnoise}}}\\
    \midrule
    \flawtag{F17}{flaw:X509TrustManagerSpecificConditions} &
    X509TrustManager Specific Conditions
    \linebreak (\opnumber{7}, \opnumber{12}, \misusenumber{16-19, 38}) &
    Insecure validation of an overridden {\scriptsize \tt checkServerTrusted} method created within an anonymous inner class created from the {\tt X509TrustManager}, due to the failure to detect {\em security exceptions thrown under impossible but \underline{context-specific} conditions, \ie conditions that seem to be relevant due to specific variable use, but are actually not};
    \eg{} {\inline{if (!(null != s || s.equalsIgnoreCase("RSA") || certs.length >= 314)) {throw new CertificateException("RSA");}}}
    \\
\midrule

    \flawtag{F18}{flaw:IntHostnameVerifierSpecificCondition} &
    Interface Hostname Verifier Specific Conditions
    \linebreak (\opnumber{8}, \opnumber{12}, \misusenumber{20}) &
    Insecure analysis of vulnerable hostname verification, \ie the {\tt \scriptsize verify()} method within an anonymous inner class (constructed similarly as in {\bf F14}), due to the failure to detect {\em a \underline{context-specific} always-true condition block that returns {\tt \scriptsize true}}; \eg{} {\scriptsize \tt if(true || session.getCipherSuite().length()>=0) return true; return false;}
    \\
\midrule

    \flawtag{F19}{flaw:AbcHostnameVerifierSpecificCondition} &
    Abstract Hostname Verifier Specific Conditions
    \linebreak (\opnumber{8}, \opnumber{12}, \misusenumber{20}) &
    Insecure analysis of vulnerable hostname verification, \ie the {\tt \scriptsize verify()} method within an anonymous inner class (constructed similarly as in {\bf F15}), due to the failure to detect {\em a \underline{context-specific} always-true condition block that returns {\tt \scriptsize true}};
    \eg{}
    \inline{if(true || session.getCipherSuite().length()>=0) return true; return false;}
    \\
\bottomrule
\end{longtable}
\end{normalsize}

    \newcolumntype{Y}[1]{>{\raggedright\arraybackslash}p{#1}}

    \begin{normalsize}

    \centering
    \begin{longtable}{
        >{}l
        >{}Y{1.5in}
        >{}p{3.6in}
    }
    \caption{Descriptions of Flaws discovered by Analyzing \detectors{} in Current Iteration}
    \label{tbl:flaws_ext}\\

    \toprule
    \textbf{ID} &
 \textbf{Flaw Name}
 \linebreak (\textbf{\footnotesize Operator, Base Misuse IDs})
 &
 \textbf{Description of Flaws} \\
    \endfirsthead
    \toprule
    \textbf{ID} &
 \textbf{Flaw Name}
 \linebreak (\textbf{\footnotesize Operator, Base Misuse IDs}) &
 \textbf{Description of Flaws} \\
    \midrule
    \endhead
    \multicolumn{3}{c}{\textit{Continued on next page.}}\\
    \endfoot
    \multicolumn{3}{p{\textwidth}}{+ flaws were observed for multiple API misuse cases,
    * Certain seemingly-unrealistic flaws may be seen in or outside a \detector's ``scope'', depending on the perspective; see Section~\ref{sec:discussion} for a broader treatment of this caveat. Full Listings are in
    Appendix~\ref{app:code_snippets}.}
    \\
    \endlastfoot
    \midrule

    \multicolumn{3}{p{\textwidth}}{\textsc{\textbf{\fcvalueresoluion+}}}\\
    \midrule

    \flawtag{F20}{flaw:StaticBytesInKeystore} &
 Static Bytes in Keystore \linebreak
 (\opnumber{19}, \misusenumber{62, 63}) &
 Not detecting bytes that are created using a static source converted to bytes and passed into IV;
    \eg{}
    \inline{new IvParameterSpec(val.getBytes(),0,8);},
    such that
    \inline{byte[] val = "12345678".getBytes();}
    \\
\midrule

    \flawtag{F21}{flaw:CharArrayToString} &
 Char Array to String \linebreak
 (\opnumber{17}, \misusenumber{40, 47, 76, 78, 94}) &
 Being unable to convert an array of Char into a String object and parsing the value;
    \eg{}
    \inline{Cipher.getInstance(String.valueOf(cryptoVariable));},
    such that
    \inline{char[] v  = "DES".toCharArray();}
    \\
\midrule

    \flawtag{F22}{flaw:UnsafeValueFromSubstring} &
 Unsafe Value from Substring \linebreak
 (\opnumber{18}, \misusenumber{40, 47, 76, 78, 94}) &
 Not detecting a Substring misuse being parsed from a longer string;
    \eg{}
    \inline{Cipher.getInstance("secureParamAES".substring(11));},
    \\
\midrule

    \flawtag{F23}{flaw:ObjectSensitivity} &
 Object Sensitivity \linebreak
 (\opnumber{16}, \misusenumber{40, 47, 76, 78, 94}) &
 Not being able to differentiate between two instances of the same object type one containing a misuse and one containing a safe value;
    \eg{} \inline{String secure = new CipherPack().safe().getpropertyName();String notsecure = new CipherPack().unsafe().getpropertyName();
    securecipher = notsecure;
    Cipher v = Cipher.getInstance(secure);}
    \\
\midrule

    \flawtag{F24}{flaw:parameterBuiltFromMethodCalls} &
 Parameter Built from Method Calls \linebreak
 (\opnumber{15}, \misusenumber{40, 47, 76, 78, 94}) &
 Unable to detect the construction of a misuse String based on calls to methods that construct the misuse \eg{}
 \lstinline[stringstyle=\color{black}, keywordstyle=\color{black}]*Class T {v="AES/GCM/NoPadding"; String A(){return "D";} String B(){return "ES";} void c(){cipher = A()+B();} String gv(){return v}} Cipher.getInstance(new T().add().gv());*
\\
\midrule

    \flawtag{F25}{flaw:parameterFromNestedConditionals} &
 Parameter from Nested Conditionals \linebreak
 (\opnumber{14}, \misusenumber{40, 47, 76, 78, 94}) &
 Being unable to keep track of a variable as it is passed through a nested number of conditional statements \eg{}
 \lstinline[keywordstyle=\color{black},stringstyle=\color{black}]*{Class T {int i = 0; c = "AES/GCM/NoPadding"; void A(){ if (i == 0){if (i == 0){ if(i == 0){ c = "AES"; } else{ c = "AES/GCM/NoPadding";}}else{ c = "AES/GCM/NoPadding"; } } else{ c = "AES/GCM/NoPadding"; }} String gv(){ return c }} c.getInstance(new T().A().gv()); *
    \\
\bottomrule

    \end{longtable}
\end{normalsize}

\newcommand{\TsmallCaseParameter}                   {\no & \ye & \ye & \ye & \ye & \nl & \ye & \ye & \ye \\}

\newcommand{\TvalueInVariable}                      {\ye & \ye & \ye & \ye & \no & \nl & \ye & \pa & \pa \\}
\newcommand{\TsecureParameterReplaceInsecure}       {\no & \no & \no & \ye & \no & \nl & \no & \ye & \ye \\}
\newcommand{\TinsecureParameterReplaceInsecure}     {\no & \no & \no & \ye & \no & \nl & \no & \no & \no \\}
\newcommand{\TstringCaseTransform}                  {\no & \no & \pa & \ye & \no & \nl & \pa & \ye & \ye \\}
\newcommand{\TnoiseReplace}                         {\no & \no & \no & \ye & \no & \nl & \no & \no & \no \\}
\newcommand{\TparameterFromMethodChaining}          {\no & \ye & \no & \ye & \no & \nl & \no & \ye & \ye \\}
\newcommand{\TdeterministicByteFromCharacterConcat} {\no & \ye & \ye & \ye & \na & \nl & \ye & \nl & \no \\}
\newcommand{\TpredictableByteFromSystemAPI}         {\no & \ye & \ye & \ye & \na & \nl & \ye & \nl & \no \\}

\newcommand{\TXExtendedTrustManager}                {\no & \na & \ye & \ye & \pa & \pa & \ye & \nl & \nl \\}
\newcommand{\TXTrustManagerSubType}                 {\no & \na & \ye & \ye & \pa & \pa & \ye & \nl & \nl \\}
\newcommand{\TIntHostnameVerifier}                  {\no & \na & \ye & \ye & \ye & \na & \ye & \nl & \nl \\}
\newcommand{\TAbcHostnameVerifier}                  {\no & \na & \ye & \ye & \ye & \na & \ye & \nl & \nl \\}

\newcommand{\TXTrustManagerGenericConditions}       {\no & \na & \ye & \ye & \ye & \no & \ye & \nl & \nl \\}
\newcommand{\TIntHostnameVerifierGenericCondition}  {\no & \na & \ye & \ye & \ye & \na & \ye & \nl & \nl \\}
\newcommand{\TAbcHostnameVerifierGenericCondition}  {\no & \na & \ye & \ye & \ye & \na & \ye & \nl & \nl \\}

\newcommand{\TXTrustManagerSpecificConditions}      {\no & \na & \no & \no & \ye & \no & \no & \nl & \nl \\}
\newcommand{\TIntHostnameVerifierSpecificCondition} {\no & \na & \ye & \no & \ye & \na & \ye & \nl & \nl \\}
\newcommand{\TAbcHostnameVerifierSpecificCondition} {\no & \na & \ye & \ye & \ye & \na & \ye & \nl & \nl \\}

\begin{table}[t]
    \centering
    \caption{\small Flaws observed in \detectors{} in previous iteration}
    \vspace{-1em}
    \label{tbl:flaw_to_tools}
	\def\arraystretch{1.2}
	\begin{tabular}{
	*{11}{C{.9cm}}
	}

	\toprule
	\textbf{Class} & \textbf{ID} & \textbf{\cryptoguardshort} & \textbf{\cognicryptshort} & \textbf{\sportbugsshort} & \textbf{\xanitizershort} & \textbf{\coverityshort} & \textbf{\qashort} & \textbf{\shiftleftshort} &\textbf{\codeqlgcsshort} & \textbf{\codeqllgtmshort}\\
	\midrule

	\multirow{1}{*}{FC1}
	& \ref{flaw:smallCaseParameter}
	&	\TsmallCaseParameter
	\hline

	\multirow{8}{*}{FC2}
	& \ref{flaw:valueInVariable}
	& \TvalueInVariable
	\cline{2-11}
	& \ref{flaw:secureParameterReplaceInsecure}*
	&	\TsecureParameterReplaceInsecure
	\cline{2-11}
	& \ref{flaw:insecureParameterReplaceInsecure}*
	&	\TinsecureParameterReplaceInsecure
	\cline{2-11}
	& \ref{flaw:stringCaseTransform}*
	&	\TstringCaseTransform
	\cline{2-11}
	& \ref{flaw:noiseReplace}*
	&	\TnoiseReplace
	\cline{2-11}
	& \ref{flaw:parameterFromMethodChaining}
	&	\TparameterFromMethodChaining
	\cline{2-11}
	& \ref{flaw:deterministicByteFromCharacters}
	&	\TdeterministicByteFromCharacterConcat
	\cline{2-11}
	& \ref{flaw:predictableByteFromSystemAPI}
	&	\TpredictableByteFromSystemAPI
	\hline

	\multirow{4}{*}{FC3}
	& \ref{flaw:X509ExtendedTrustManager}
	&	\TXExtendedTrustManager
	\cline{2-11}

	& \ref{flaw:X509TrustManagerSubType}
	&	\TXTrustManagerSubType
	\cline{2-11}

	& \ref{flaw:IntHostnameVerifier}
	&	\TIntHostnameVerifier
	\cline{2-11}

	& \ref{flaw:AbcHostnameVerifier}
	&	\TAbcHostnameVerifier
	\hline

	\multirow{3}{*}{FC4}
	& \ref{flaw:X509TrustManagerGenericConditions}
	&	\TXTrustManagerGenericConditions
	\cline{2-11}
	& \ref{flaw:IntHostnameVerifierGenericCondition}
	&	\TIntHostnameVerifierGenericCondition
	\cline{2-11}
	& \ref{flaw:AbcHostnameVerifierGenericCondition}
	&	\TAbcHostnameVerifierGenericCondition
	\hline

	\multirow{3}{*}{FC5}
	& \ref{flaw:X509TrustManagerSpecificConditions}
	&	\TXTrustManagerSpecificConditions
	\cline{2-11}
	& \ref{flaw:IntHostnameVerifierSpecificCondition}
	&	\TIntHostnameVerifierSpecificCondition
	\cline{2-11}
	& \ref{flaw:AbcHostnameVerifierSpecificCondition}
	&	\TAbcHostnameVerifierSpecificCondition
	\bottomrule
	\multicolumn{11}{p{.98\textwidth}}
	{
		\no{} = Flaw Present, \ye{} = Flaw Absent, \pa{} = Flaw partially present, \na = detector does not claim to handle the misuse associated with the flaw, \nl= detector claims to handle but did not detect base version of misuse; CG~=~CryptoGuard,  CC~=~\cognicrypt, SB~=~\spotbug, \xanitizershort~=~\xanitizer, \coverityshort~=~\coverity, QA~=~\qark, \shiftleftshort~=~ShiftLeft, \codeqlgcsshort~=~\codeqlgcs.
	}\\
	\multicolumn{11}{p{.98\textwidth}}
	{
		*Certain seemingly-unrealistic flaws may be seen in/outside a \detector's ``scope'', depending on the perspective; see Section~\ref{sec:discussion} for a broader treatment of this caveat.
	}
\end{tabular}
\end{table}

\newcommand{\nTsmallCaseParameter}                   { \nye & \ye & \ye & \nl & \ye & \ye  & \ye & \ye & \ye & \pa & \nl & \pr \\}
\newcommand{\nTvalueInVariable}                      { \ye  & \ye & \ye & \nl & \ye & \pa  & \pa & \ye & \pa & \ye & \nl & \ye \\}
\newcommand{\nTsecureParameterReplaceInsecure}       { \nye & \no & \no & \nl & \no & \ye  & \ye & \no & \no & \no & \nl & \no \\}
\newcommand{\nTinsecureParameterReplaceInsecure}     { \nye & \no & \no & \nl & \no & \nye  & \no & \ye & \no & \no & \nl & \no \\}
\newcommand{\nTstringCaseTransform}                  { \nye & \no & \pa & \nl & \pa & \ye  & \ye & \no & \no & \no & \nl & \no \\}
\newcommand{\nTnoiseReplace}                         { \no  & \no & \no & \nl & \no & \nye  & \no & \ye & \no & \no & \nl & \no \\}
\newcommand{\nTparameterFromMethodChaining}          { \no  & \ye & \no & \nl & \no & \ye  & \ye & \na & \no & \no & \nl & \no \\}
\newcommand{\nTdeterministicByteFromCharacterConcat} { \no  & \ye & \ye & \nl & \ye & \nl  & \nl & \na & \ye & \no & \nl & \nl \\}
\newcommand{\nTpredictableByteFromSystemAPI}         { \no  & \ye & \ye & \nl & \ye & \nl  & \nl & \nl & \ye & \no & \nl & \nl \\}
\newcommand{\nTXExtendedTrustManager}                { \no  & \na & \ye & \pr & \ye & \no  & \no & \nl & \ye & \ye & \nl & \ye \\}
\newcommand{\nTXTrustManagerSubType}                 { \no  & \na & \ye & \pr & \ye & \no  & \no & \nl & \ye & \ye & \nl & \ye \\}
\newcommand{\nTIntHostnameVerifier}                  { \no  & \na & \ye & \na & \ye & \nl  & \nl & \nl & \ye & \ye & \nl & \ye \\}
\newcommand{\nTAbcHostnameVerifier}                  { \no  & \na & \ye & \na & \ye & \nl  & \nl & \nl & \ye & \ye & \nl & \ye \\}
\newcommand{\nTXTrustManagerGenericConditions}       { \no  & \na & \ye & \no & \ye & \no  & \no & \nl & \no & \no & \nl & \no \\}
\newcommand{\nTIntHostnameVerifierGenericCondition}  { \no  & \na & \ye & \na & \ye & \nl  & \nl & \nl & \no & \no & \nl & \no \\}
\newcommand{\nTAbcHostnameVerifierGenericCondition}  { \no  & \na & \ye & \na & \ye & \nl  & \nl & \nl & \no & \no & \nl & \no \\}
\newcommand{\nTXTrustManagerSpecificConditions}      { \no  & \na & \no & \no & \no & \no  & \no & \nl & \no & \no & \nl & \no \\}
\newcommand{\nTIntHostnameVerifierSpecificCondition} { \no  & \na & \ye & \na & \ye & \nl  & \nl & \nl & \no & \no & \nl & \no \\}
\newcommand{\nTAbcHostnameVerifierSpecificCondition} { \no  & \na & \ye & \nl & \ye & \nl  & \nl & \nl & \no & \no & \nl & \no \\}
\newcommand{\nStaticBytesInKeystore}                 { \ye  & \ye & \ye & \nl & \ye & \nl  & \nl & \nl & \ye & \ye & \nl & \nl \\}
\newcommand{\nCharArrayToString}                     { \ye  & \no & \no & \nl & \no & \ye  & \no & \no & \no & \no & \nl & \no \\}
\newcommand{\nUnsafeValueFromSubstring}              { \no  & \no & \no & \nl & \no & \ye  & \no & \no & \no & \no & \nl & \no \\}
\newcommand{\nObjectSensitivity}                     { \no  & \no & \no & \nl & \no & \no  & \no & \no & \no & \no & \nl & \no \\}
\newcommand{\nparameterBuiltFromMethodCalls}         { \no  & \no & \no & \nl & \no & \no  & \no & \no & \no & \no & \nl & \no \\}
\newcommand{\nparameterFromNestedConditionals}       { \no  & \no & \no & \nl & \no & \no  & \no & \no & \no & \no & \nl & \no \\}
\newcommand{\nColumnNames}                           { nCG  & nCC & nSB & nQA & nSL & nGCS & nLG & AC  & SQ  &  SY & CD  & DS  \\}

\begin{table}[tbp]
    \centering
	\normalsize
    \caption{\small Flaws observed in \detectors in current iteration}
	\setlength{\tabcolsep}{2pt}
    \label{tbl:new_flaw_to_tools}
	\begin{tabularx}{\textwidth}{
        c C{1cm} *{12}{X}
    }
	\toprule
	\textbf{Class} & \textbf{ID}  &
	\multicolumn{7}{>{}c}{New versions of \Detectors{}} &
	\multicolumn{5}{>{}c}{Newly introduced \Detectors{}}
	\\
	\cmidrule(lr){3-9}\cmidrule(lr){10-14}
	\textbf{} &  &
	\textbf{\newcryptoguardshort} &
	\textbf{\newcognicryptshort} &
	\textbf{\newspotbugsshort} &
	\textbf{\newqarkshort} &
	\textbf{\newshiftleftshort} &
	\textbf{\newlgtmshort{}} &
	\textbf{\newcqversion} &
	\textbf{\codegurushort} &
	\textbf{\sonarqubeshort} &
	\textbf{\snykshort} &
	\textbf{\codigashort} &
	\textbf{\deepsourceshort}
	\\
	\midrule

	\multirow{1}{*}{FC1}
	& F1
	&	\nTsmallCaseParameter
	\hline

	\multirow{8}{*}{FC2}
	& F2
	& \nTvalueInVariable
	\cline{2-14}
	& F3*
	&	\nTsecureParameterReplaceInsecure
	\cline{2-14}
	& F4*
	&	\nTinsecureParameterReplaceInsecure
	\cline{2-14}
	& F5*
	&	\nTstringCaseTransform
	\cline{2-14}
	& F6*
	&	\nTnoiseReplace
	\cline{2-14}
	& F7
	&	\nTparameterFromMethodChaining
	\cline{2-14}
	& F8
	&	\nTdeterministicByteFromCharacterConcat
	\cline{2-14}
	& F9
	&	\nTpredictableByteFromSystemAPI

	\cline{2-14}
	& F20
	&	\nStaticBytesInKeystore
	\cline{2-14}
	& F21
	&	\nCharArrayToString
	\cline{2-14}
	& F22
	&	\nUnsafeValueFromSubstring
	\cline{2-14}
	& F23
	&	\nObjectSensitivity
	\cline{2-14}
	& F24
	&	\nparameterBuiltFromMethodCalls
	\cline{2-14}
	& F25
	&	\nparameterFromNestedConditionals

	\hline

	\multirow{4}{*}{FC3}
	& F10
	&	\nTXExtendedTrustManager
	\cline{2-14}

	& F11
	&	\nTXTrustManagerSubType
	\cline{2-14}

	& F12
	&	\nTIntHostnameVerifier
	\cline{2-14}

	& F13
	&	\nTAbcHostnameVerifier
	\hline

	\multirow{3}{*}{FC4}
	& F14
	&\nTXTrustManagerGenericConditions
	\cline{2-14}
	& F15
	&\nTIntHostnameVerifierGenericCondition
	\cline{2-14}
	& F16
	&	\nTAbcHostnameVerifierGenericCondition
	\hline

	\multirow{3}{*}{FC5}
	& F17
	&	\nTXTrustManagerSpecificConditions
	\cline{2-14}
	& F18
	&	\nTIntHostnameVerifierSpecificCondition
	\cline{2-14}
	& F19
	&	\nTAbcHostnameVerifierSpecificCondition

	\bottomrule
	\multicolumn{14}{>{\raggedright\arraybackslash}p{\textwidth}}
	{
		\no{} = Flaw Present, \ye{} = Flaw Absent, \nye = Flaw was present in the previous study, is no longer present after responsible disclosure \pr{} = Flaw partially present, \na = detector does not claim to handle the misuse associated with the flaw, \np = detector claims to handle but did not detect base version of misuse \codegurushort~=~\codeguru, \sonarqubeshort~=~\sonarqube, \snykshort~=~\snyk, \codigashort~=~\codiga, \deepsourceshort~=~\deepsource, \newcryptoguardshort~=~\cryptoguardupdate, \newcognicryptshort~=~\cognicryptupdate, \newspotbugsshort~=~\spotbugsupdate, \newqarkshort~=~\qarkupdate, \newshiftleftshort~=~\shiftleftupdate, n\codeqlgcsshort~=~\codeqlversion, nLGTM~=~\newlgtm.
		}\\
	\multicolumn{14}{>{\raggedright\arraybackslash}p{\textwidth}}
	{
		*Certain seemingly-unrealistic flaws may be seen in/outside a \detector's ``scope'', depending on the perspective; see Discussion for a broader treatment of this caveat.
	}
\end{tabularx}
\vspace{-2em}
\end{table}

We organize these flaws into \countFlawClassesText{} \textit{flaw classes} (\ref{rq:classes}), representing the shared limitations that caused them.
Table~\ref{tbl:flaws} and Table~\ref{tbl:flaws_ext} provides the complete list of the flaws, categorized along flaw classes, while Table~\ref{tbl:flaw_to_tools} and Table~\ref{tbl:new_flaw_to_tools} provides a mapping of the flaws to the \detectors that exhibit them.

As we have shown in Table~\ref{tbl:flaw_to_tools} in our previous study, a majority of the total flaws (computed by adding \ \no\ and\ \nl\ instances) identified in \detectors, \ie 45/76 or 59.21\% can be {\em solely} attributed to our mutation-based approach, whereas only 31/76, \ie 40.79\% could {\em also} be found using base instantiations of the corresponding misuse cases.
Further, all flaws in 6/9 \detectors were only identified using \tool.
This demonstrates the advantage of using mutation, over simply evaluating with base instantiations of misuses from the taxonomy.
\codeqlgcsshort, \codeqllgtmshort, and \qark fail to detect base cases due to incomplete rule sets (see
online Appendix~\cite{masc-online}
 for discussion).
Similarly, in our current study, a significant portion of the total flaws (122/196) we identified in \detectors can be attributed to mutation based approach, whereas
74 could be found using base instantiations of the corresponding misuse cases as shown in Table~\ref{tbl:new_flaw_to_tools}.
Moreover, 6 fixed flaws (\nye in Table~\ref{tbl:new_flaw_to_tools}) in this extended study, \ie flaws that were present in our previous study, but are no longer present, can be attributed to \tool{} because of the responsibly disclosed reports submitted in \oakland{}.

In our previous study, at the initial stage of our evaluation (\ie before we analyzed uncaught mutants), we discovered that certain \detectors~\cite{rxa+19, KSA+18} analyze only a limited portion of the target applications, which greatly affects the reliability of their results.
As these gaps were not detected using \tools evaluation, we do not count these in our flaws or flaw classes.
However, due to their impact on the basic guarantees provided by \detectors, we believe that such gaps must be discussed, addressed, and avoided by future \detectors, which is why we discuss them under a flaw class {\em zero}.

\myparagraphnew{Flaw Class Zero (FC0) -- {\em Incomplete analysis of target code}}
Starting from Android API level 21 (Android 5.0), apps with over 64k reference methods are automatically split to multiple Dalvik Executable (DEX) byte code files in the form of \texttt{\small classes<N>.dex}.
Most apps built after 2014 would fall into this category, as the Android Studio IDE automatically uses {\em multidex} packaging for any app built for Android 5.0 or higher~\cite{EnableMultidexApps}.
However, we discovered that \cryptoguard and \cognicrypt do not handle multiple dex files ~(see~Table~\ref{tbl:flaw_to_tools}), despite being {\em released 4 and 5 years after Android 21, respectively}. %
Given that this flaw affected \cryptoguard's ability to analyze a majority of Android mutants (\ie detecting only $871/2515$),
 we developed a patch that fixes this issue, and used the patched version of \cryptoguard for our evaluation. %
The \cognicrypt maintainers confirmed that they are working on this issue in their upcoming release, which is why we did not create a patch, but used non-multidex minimal examples to confirm that each flaw discovered in \cognicrypt is not due to the multidex issue.
Similarly, we discovered that \cryptoguard ignores any class whose package name contains ``{\tt android.}'' or ends in ``{\tt android}'', which prevents it from analyzing important apps such as LastPass (\texttt{com.lastpass.lpandroid}) and LinkedIn (\texttt{com.linkedin.android}), which indicates the severity of even trivial implementation flaws.
 We submitted a patch to address this flaw and evaluated the patched version~\cite{mascArtifact}.%
Finally, \codiga{} was unable to detect any of the crypto-API misuse we analyzed, even though it explicitly claims that it \textit{"supports hundreds of rules for Java, checking that your code is safe and secure"}~\cite{codiga_description}. When we reached out to Codiga (April 2023), we were informed that Codiga would shut down by May 3rd, 2023 without addressing this issue.
As a result, we have marked all flaws accordingly (\np{}), \ie claims to handle but does not detect base version of misuse in Table~\ref{tbl:new_flaw_to_tools}.

The remainder of this section discusses each flaw class with representative examples as they manifest in \detectors and the {\em impact} of the flaws in terms of examples found in real software.

\myparagraphnew{FC1: String Case Mishandling (F1)}
As discussed in the motivating example (and seen in {\bf \ref{flaw:smallCaseParameter}} in Table~\ref{tbl:flaws}), a developer may use \texttt{des} or \texttt{dEs} (instead of \DES) in \ciphergetinstance{} without JCA raising exceptions.
As shown in Table~\ref{tbl:flaw_to_tools}, \cryptoguard{} did not detect such misuse in our previous study.
We submitted a patch to \cryptoguard to address this flaw, which was accepted, and demonstrates that this flaw was recognized as an \underline{\em implementation gap} by the \cryptoguard developers~\cite{mascArtifact}.
Since the latest version of \cryptoguard{} can identify this misuse, we have marked it as such (\nye{}) in Table~\ref{tbl:new_flaw_to_tools}.
\snyk{} is able to detect when the misuse is instantiated using the \messageDigestInstance{} method (using "md5"), but not using the \ciphergetinstance{} method (using "aes", which defaults to ECB mode). Similarly, \deepsource could detect it for the \ciphergetinstance{} method, but not \messageDigestInstance{} method, indicating a partial \underline{\em implementation} gap in both these \detectors{}.

\myparagraphnew{FC2: Incorrect Value Resolution (F2 -- F9, F20 -- F25)}
The flaws in this class occur due to the inability of $13$/$14$ of the \detectors to resolve parameters passed to crypto-APIs.
For example, consider {\bf \ref{flaw:valueInVariable}}, previously discussed in Listing~\ref{lst:des-example} in Section~\ref{sec:introduction}, where the string value of an algorithm type (\eg \DES) is passed through a variable to \ciphergetinstance.
\coverity was not able to detect this (mis)use instance, hence exhibiting {\bf \ref{flaw:valueInVariable}} (Table~\ref{tbl:flaw_to_tools}), and in fact, demonstrated a consistent inability to resolve parameter values (flaws {\bf \ref{flaw:valueInVariable}} -- {\bf \ref{flaw:parameterFromMethodChaining}}), indicating a \underline{\em design gap}.
On the contrary, as \spotbug partially detects the type of misuse represented in {\bf \ref{flaw:stringCaseTransform}}, \ie when it is instantiated using the \ciphergetinstance{}method, but not using the \messageDigestInstance{} method, which indicates an \underline{\em implementation gap}, which continues to exist throughout the previous and current study.

Further, LGTM and GCS are partially susceptible to \ref{flaw:valueInVariable} because of an intricate problem in their rule-sets.
That is, both tools are capable of tracking values passed from variables, and generally detect mutations similar to the one in \ref{flaw:valueInVariable} (\ie and also the one in Listing~\ref{lst:des-example}, created using \opnumber{2}).
However, one of the mutant instances that we created using \opnumber{2} used \AES in \ciphergetinstance, which may seem correct but is actually a misuse, since specifying \AES alone initializes the cipher to use the highly vulnerable \ECB mode as the mode of operation. %
Unfortunately, both LGTM and GCS use mostly similar CodeQL rule-set~\cite{codeql_basic_crypto_rules} which does not consider this nuance, leading both tools to ignore this misuse.

Interestingly, \sonarqube{}, as shown in Table~\ref{tbl:new_flaw_to_tools}, is partially susceptible to \ref{flaw:valueInVariable}, as it does detect misuse for \ciphergetinstance{} method, but not for \messageDigestInstance{} because of an entirely different reason.
As we found from the discussion with \sonarqube{} during the responsible disclosure, this flaw stems from the tendency of avoiding false-positives.
As they elaborated, evaluating a non-constant value may result in increased number of false-positives, and as a result, the analyzer avoids evaluating non-constant identifiers in this particular case.
While weak hashing algorithm and weak encryption are within OWASP's TOP 10 crypto API issues and are considered within scope by \sonarqube{}, this reveals that developers of the same \detector{} may treat crypto-APIs differently based on internal prioritization factors, which led to this particular flaw. We address this particular insight in Section~\ref{sec:discussion} later on.

Finally, we observe that \cognicrypt detects some of the more complex behaviors represented by flaws in {\bf FC2} (\ie\  {\bf \ref{flaw:parameterFromMethodChaining}} -- {\bf \ref{flaw:predictableByteFromSystemAPI}}), but does not detect the simpler ones ({\bf \ref{flaw:secureParameterReplaceInsecure}} -- {\bf \ref{flaw:noiseReplace}}).
From our interaction with \cognicrypt's maintainers, we have discovered that \cognicrypt should be able to detect such transformations by design, as they deviate from \crysl rules.
However, in practice, CogniCrypt cannot reason about certain transformations at present (but could be modified to do so in the future), and produces an ambiguous output that neither flags such instances as misuse, nor as warnings for manual inspection, due to an \underline{implementation gap}.
The developers agree that CogniCrypt should clearly flag the API-use that it does not handle in the report, and refer such use to manual inspection.

\noindent
\myparagraphnew{Impact ({FC2})} We found misuse similar to the instance in {\bf F2} in Apache Druid, an app with $10.3K$ stars and $400$ contributors on Github (see Listing~\ref{lst:transformationstring:apache:druid} in
Appendix~\ref{app:code_snippets})
.
Further, we found real apps that convert the case of algorithm values before using them in a restrictive crypto API~\cite{realMisuseAlgoUpperCase} ({\bf \ref{flaw:stringCaseTransform}}, instantiated using \opnumber{3}), or process values to replace "noise"~\cite{realMisuseNoiseReplace} ({\bf F3}, {\bf F4}, {\bf F6}, instantiated using \opnumber{4}).
We observed that ExoPlayer, a media player from Google with over $16.8$K stars on GitHub, used the predictable {\em Random} API for creating \ivparameterspec objects~\cite{realMisuseRandomIV} until 2019, similar in nature to {\bf \ref{flaw:predictableByteFromSystemAPI}}.
Developers also use constants for IVs ({\bf \ref{flaw:deterministicByteFromCharacters}}), as seen in UltimateAndroid~\cite{realMisuseConstantIVUltimateAndroidGithub} ($2.1$K stars), and JeeSuite ($570$ stars)~\cite{realMisuseConstantIVJeeSuiteAndroidGithub}.

We also found instances of these flaws in apps that were analyzed with a \detector, specifically, LGTM.
Apache Ignite~\cite{apache_ignite} ($360$ contributors, $3.5$K stars) contains a misuse instance similar to one that led to {\bf F2}, where only the name of the cipher is passed to the \ciphergetinstance{} API~\cite{apache_ignite_key_get_instance} which causes it to default to ``ECB'' mode.
\codeqllgtmshort{} does not report this as it considers \ECB use as insecure for Java (but oddly secure for JavaScript).
We found similar instances of \ECB misuse in Apache-Hive~\cite{aes_apache_hive} ($250$ contributors, $3.4$K stars), Azure SDK for Java~\cite{azure_sdk_java} ($328$ contributors, $857$ stars), which LGTM would not detect.
Finally, in Apache Ignite~\cite{apache_ignite}, we found a \ciphergetinstance{} invocation that contained a method call in place of the cipher argument (\ie a chain of length 1, a basic instance of {\bf F7})~\cite{apache_ignite_key_get_instance}.

\myparagraphnew{FC3: Incorrect Resolution of Complex Inheritance and Anonymous Objects (\ref{flaw:X509ExtendedTrustManager} -- \ref{flaw:AbcHostnameVerifier})}
The flaws in this class occur due to the inability of \detectors to resolve complex inheritance relationships among classes, generally resulting from applying \textit{flexible} mutation operators (Sec.~\ref{sec:flexible-ops}) to certain misuse cases.
For example, consider {\bf \ref{flaw:X509TrustManagerSubType}} in Table~\ref{tbl:flaws}, also illustrated in Listing~\ref{lst:aic_x509tm} in
Appendix~\ref{app:code_snippets}
.
Further, we find that several \detectors, \eg \xanitizer, \spotbug and \sonarqube are immune to these flaws, which indicates that traversing intricate inheritance relationships is a {\em design consideration for some \detectors}, and a \underline{\em design gap} in others such as \cryptoguard and \qark.
Moreover, such indirect relationships can not only be expected from evasive developers (\ie \ref{adversary:evasive}) but is also found in real apps investigated by the \detectors, as described below.

\noindent
\myparagraphnew{Impact (FC3)}
We found an exact instance of the misuse representing {\bf F10} (generated from {\bf OP$_{12}$}) in the class \inlinesmall{TrustAllSSLSocketFactory} in Apache JMeter~\cite{realMisuseX509ExtendedTrustManager} ($4.7$K stars in GitHub).
{\bf F11} is the generic version of {\bf F10}, and fairly common in repositories and libraries (\eg BouncyCastle~\cite{BCX509ExtendedTrustManager, FilterX509TrustManager}).
{\bf F12} and {\bf F13} were also generated using {\bf OP$_{12}$}, but with the \inlinesmall{HostnameVerifier}-related misuse, and we did not find similar instances in the wild in our limited search.

\myparagraphnew{FC4: Insufficient Analysis of Generic Conditions in Extensible Crypto-APIs (\ref{flaw:X509TrustManagerGenericConditions} -- \ref{flaw:AbcHostnameVerifierGenericCondition})}
The flaws in this class represent the inability of certain \detectors to identify fake conditions within overridden methods, \ie unrealistic conditions, or always true condition blocks (\eg Listing~\ref{lst:xtrustManagerGenericConditions} in
Appendix~\ref{app:code_snippets} shows for {\bf \ref{flaw:X509TrustManagerGenericConditions}}).
Flaws in this class represent the behavior of an evasive developer (\ref{adversary:evasive}). Several \detectors{}, \eg
\xanitizer and \spotbug, can identify such spurious conditions.

\myparagraphnew{FC5: Insufficient Analysis of Context-specific Conditions in Extensible Crypto-APIs (\ref{flaw:X509TrustManagerSpecificConditions} -- \ref{flaw:AbcHostnameVerifierSpecificCondition})}
The flaws in this class represent misuse similar to {\bf FC4}, except that the fake conditions used here are contextualized to the overridden function, \ie they check {\em context-specific} attributes (\eg the length of the certificate chain passed into the method,  {\bf \ref{flaw:X509TrustManagerSpecificConditions}}).
An evasive developer may attempt this to add further realism to fake conditions to evade tools such as \xanitizer that are capable of detecting generic conditions.
Indeed,
we observe that \xanitizer fails to detect misuse when context-specific conditions are used, for both {\bf \ref{flaw:X509TrustManagerSpecificConditions}} and {\bf \ref{flaw:IntHostnameVerifierSpecificCondition}}.
Our suspicion is that this weakness is due to an optimization, which exempts conditions from analysis if they seem realistic.%

Particularly, we observe that \xanitizer correctly detects the fake condition in {\bf \ref{flaw:AbcHostnameVerifierSpecificCondition}}, and that the only difference between {\bf \ref{flaw:AbcHostnameVerifierSpecificCondition}} and {\bf \ref{flaw:IntHostnameVerifierSpecificCondition}} is that the instances of misuse they represent occur under slightly different class hierarchies.
Hence, our speculation is that this an {\em accidental benefit}, \ie the difference could be the result of an {\em incomplete implementation of the \underline{unnecessary optimization}} across different class hierarchies.
\spotbugfull shows a similar trend, potentially because \xanitizer uses \spotbug for several SSL-related analyses.
Finally, we observe that \coverity is immune to both generic {\bf FC4}) and context-specific fake conditions {\bf FC5}).

\noindent
\myparagraphnew{Impact (FC4, FC5)}
In this Stack Overflow post~\cite{realMisuseStackCondition}, the developer describes several ways in which they tried to get Google Play to accept their faulty \trustManager implementation, one of which is exactly the same as the misuse instance that led to {\bf F17} (generated using \opnumber{7} and \opnumber{12}), which is a more specific variant of {\bf F14} (generated \opnumber{7}, \opnumber{9} and \opnumber{12}), as illustrated in Listing~\ref{lst:specific_condition_trustmanager} in
Appendix~\ref{app:code_snippets}
.
We observe similar evasive attempts towards vulnerable hostname verification~\cite{realMisuseHostnameVerifierGenericCondition} which are similar in nature to {\bf F15} and {\bf F16}, and could be instantiated using \opnumber{8} and \opnumber{10}.
We also found developers trying to evade detection by applying context-specific conditions in the hostname verifier~\cite{realMisuseHostnameVerifierSpecificCondition}, similar to {\bf F18} and {\bf F19}.

\section{Limitations}\label{sec:limitations}
\tool{} does not attempt  to replace formal verification, and hence, does not guarantee that all flaws in a \detector will be found.
Instead, it enables systematic evaluation of \detectors, which is an advancement over manually curated benchmarks.
Aside from this general design-choice, our approach has the following limitations:

\myparagraphnew{1. Completeness of the Taxonomy} To ensure a taxonomy that is as comprehensive as possible, we meticulously follow best-practices learned from prior work~\cite{KBB+09,CorbinStrauss}, and also ensure labeling by two authors.
However, the fact remains that regardless of how systematic our approach is, due to the manual and generalized of the SLR, we may miss certain subtle contexts during the information extraction phase (see specific examples in
Online Appendix~\cite{masc-online}
). %
Thus, while not necessarily complete, this taxonomy, generated through over {\em 2 person months of manual effort}, is, to the best of our knowledge, the most comprehensive in recent literature.

\myparagraphnew{2. Focus on Generic Mutation Operators} We have currently constructed generic operators based on typical {\em usage conventions}, \ie to apply to as many misuse instances from the taxonomy as possible.
However, currently, \tool does not incorporate operators that may fall outside of usage-based conventions, \ie which may be more tightly coupled with specific misuse cases, such as an operator for calling \inlinesmall{create} and \inlinesmall{clearPassword} in a specific order for \inlinesmall{PBEKeySpec}.
We plan to incorporate such operators into \tool in the future.

\myparagraphnew{3. Focus on Java and JCA} \tools{} approach is largely informed by JCA and Java. Additional operators and adjustments will be required to adapt \tool{}'s implementation to JCA-equivalent frameworks in Java, specially when adapting our usage-based mutation operators to non-JCA conventions. For languages that support the OOP paradigm, such as CPP and Go, we assume \tool{}'s approach will remain the same as frameworks tend to create and maintain consistent patterns specific to languages. Therefore, the implementation of \tool{} will require translating the existing operators and scopes to equivalent OOP-based structures in the context of those languages. %

\myparagraphnew{4. Evolution of APIs} Future, tangential changes in how JCA operates might require changing the implementation of \tools{} mutation operators.
Furthermore, incremental effort will be required to incorporate new misuse cases that are discovered with the evolution of crypto APIs.
We have designed \tool{} to be as flexible as possible by means of reflection and automated code generation for mutation operators, which should make adapting to such changes easier.

\myparagraphnew{5. Relative Effectiveness of Individual Operators}
This paper demonstrates the key claims of MASC, and its overall effectiveness at finding flaws, but does not evaluate/claim the {\em relative usefulness of each operator} individually. A comprehensive investigation of relative usefulness would require the mutation of all/most misuse cases from the taxonomy, with every possible operator and scope, a broader set of apps to mutate, and a complete set of crypto-detectors, which is outside the scope of \tool, but a direction for future work.

\section{Discussion and Conclusion}
\label{sec:discussion}

Designing \detectors is in no way a simple task; as these have to balance several orthogonal requirements, such as detecting as many vulnerabilities as possible, without introducing false positives, while scaling to large codebases. %
Yet, the fact is that there is significant room for improvement in how \detectors are built and evaluated, as evidenced by the flaws discovered using \tool.

To move forward, we need to understand the divergent perspectives regarding the design of \detectors, and reach a consensus (or at least an agreeable minima) in terms of what is expected from \detectors and how we will design and evaluate them to satisfy the expectations. %
We seek to begin this discourse within the security community by integrating several views on the design decisions behind \detectors, informed by our results and conversations with tool designers (quoted with consent) during the vulnerability reporting process.%

\subsection{Security-centric Evaluation vs. Technique-centric Design}
Determining what misuse is within or outside the scope for a \detector is a complex question that yields several different viewpoints.
This paper's view is {\em security-centric}, \ie even if some misuse instances may seem unlikely or evasive, \detectors that claim/target security-focused use cases (\eg compliance, auditing) should account for them.
However, we observe that tool designers typically adhere to a {\em technique-centric} perspective, \ie the design of \detectors is not influenced by a threat model, but mainly by what static analysis can and cannot accomplish while implicitly assuming a benign developer.
This quote from the maintainers of \cryptoguard highlights this view, wherein they state that the ``lines'' between what is within/outside scope \emphdblquote{seen so far were technically motivated -- not use-case motivated..should we use alias analysis?...}.
This gap in perspective does not mean that \detectors may not detect any of the mutants generated by \tool using operators based on the \ref{adversary:evasive} threat model; rather, it only means that detection (or lack thereof) may not be {\em caused} by a security-centric design.

\subsection{Defining ``Scope'' for the Technique-centric Design}
\label{sec:discussion-within-scope}
We observe that even within \detectors that take a \textit{technique-centric} approach, there is little agreement on the appropriate scope of detection.
For instance, \xanitizer focuses on catching every possible misuse instance, regardless of any external factors such as whether that kind of misuse is observed in the wild, or a threat model, as the designers believe that \emphdblquote{the distinction should not be between `common' and `uncommon', but instead between `can be (easily) computed statically' and `can not be computed'.}.
This makes it possible for \xanitizer to detect unknown or rare problems, but may also result in it not detecting a commonly observed misuse that is hard to compute statically, although we did not observe such cases. For \sonarqube{}, the "cost" of analyzing a pattern versus whether it is easy to detect through manual analysis is an additional consideration,
``\textit{... easy to identify for simple human readers. For our analyzer ... this kind of method is expensive ... issue is more of a misconfiguration on the developer's side, trying to cover this type of case seems a bit out of scope.}''

In contrast, \cryptoguard, \cognicrypt, and GCS/\codeqllgtm (same developers) would {\em consider seemingly\-unlikely/evasive flaws} {\em within scope} (\eg\ \fnumber{3} -- \fnumber{6}, \fnumber{8}), because they were found in the wild (unlike \xanitizer, for which this is not a consideration).
This view aligns with our perspective, that regardless of how it was generated, if a misuse instance (representing a flaw) is discovered in real apps (which is {\em true for all flaws except \fnumber{12} and \fnumber{13}}, it should be within the detection scope.
However, GCS/\codeqllgtm maintainers extend this definition with the condition that the observations in the wild be frequent, to motivate change.
Note that this scope defining factor also extends to the additional, extended security test--suite of \codeqlgcs, while the default test--suite focuses on precision, as the number of false-positives is a concern for developers of \detectors{}, likely inherited from the program analysis community studies~\cite{JSMB13, IRFW19, CB16}.
These divergent perspectives motivate the need to clearly define the expectations from \detectors
by the designers, \ie \detectors{} need to clearly communicate if their use cases are for hostile/adversarial context (\eg security audit, certification, or compliance), or for developer-friendly, \textit{helping-with-coding} context. Furthermore, to avoid creating a false sense of security, it is important to document known limitations, such as design considerations, and share them with (potential) users, so that the users can make informed decisions about their capabilities.

\subsection{Utility of Seemingly-Uncommon or Evasive Tests}
\label{sec:discussion-evasive}
As Bessey et al. state from practical deployment experience in 2010~\cite{BBC+10}, \emphdblquote{No bug is too foolish to check for}, and that \emphdblquote{Given enough code, developers will write almost anything you can think of...}.
The results from our evaluation and the impact study corroborate this sentiment, \ie\ \fnumber{3} -- \fnumber{6} and \fnumber{8} were all obtained using operators (\opnumber{3}, \opnumber{4}, and \opnumber{6}) modeled to emulate threats \ref{adversary:benign_accident} and \ref{adversary:benign_fixing}, \ie representing {\em benign} behavior (however unlikely); and indeed, these flaws were later found in supposedly benign applications.
This suggests that the experience of Bessey et al. is valid a decade later, making it important to evaluate \detectors with ``more-than-trivial'' cases to not only test their detection prowess, but to also account for real problems that may exist in the wild.

\subsection{The Need to Strengthen Crypto-Detectors}
\label{sec:discussion-need-stronger}
We argue that it is not only justified for tools to detect uncommon cases (\eg given that even benign developers write seemingly-unlikely code), but also critical for their sustained relevance.
As designers of Coverity found~\cite{BBC+10}, false negatives matter from a commercial perspective, because \emphdblquote{Potential customers intentionally introduced bugs into the system, asking {`Why didn't you find it?'}}.

Perhaps more significantly, the importance of automated \detectors with the ability to guarantee {\em assurance} is rising with the advent of new {\em compliance} legislation such as the IoT Cybersecurity Improvement Act of 2020~\cite{iot-bill}, which seeks to rein in vulnerabilities in billions of IoT systems that include vulnerable server-side/mobile components.
Vulnerabilities found {\em after} a compliance certification may result in penalties for the developers, and financial consequences for the compliance checkers/labs and \detectors used. %
Complementing static detection with manual or dynamic analysis may be infeasible at this scale, as tool designers noted:
\eg\ \emphdblquote{...review an entire codebase at once, manual review can be difficult.} (\codeqllgtm) and \emphdblquote{Existing dynamic analysis tools will be able to detect them only if the code is triggered (which can be notoriously difficult)} (\cryptoguard).
Thus, static \detectors will need to become more robust, and capable of detecting hard-to-detect misuse instances.

\subsection{Towards Crypto-Detectors Strengthened by a Security-Centric Evaluation}
\label{sec:evolution-conclusion}

Fortunately, we observe that there is support among tool designers for moving towards stronger security guarantees. %
For instance, \cognicrypt designers see a future research direction in expressing evasive scenarios in the \crysl language \ie \emphdblquote{...what would be a nice future direction is to tweak the SAST with such optimizations/ more analysis but still allow the \crysl developer to decide if he wants to switch these `evasive user' checks...},  but indicate the caveat that developers may not use such additional configuration options~\cite{JSMB13}.
However, we believe that such options will benefit {\em independent evaluators}, and developers who are unsure of the quality of their own supply chain, to perform a hostile review at their disposal.
Similarly, the \cryptoguard designers state that \emphdblquote{...this insight of evasive developers is very interesting and timely, which immediately opens up new directions.}

This potential paradigm-shift towards a {\em security-focused} design of \detectors is timely, and
\tool's rigorous evaluation with expressive test cases can play an effective role in it, by enabling tool designers to proactively address gaps in detection during the design phase itself, rather than reactively (\ie after a vulnerability is discovered in the wild).
More importantly, further large-scale evaluations using \tool, and the flaws discovered therein, will enable the community to continually examine the design choices made by \detectors and reach consensus on what assurances we can feasibly expect. %
We envision that such development aided by \tool will lead to a mature ecosystem of \detectors with a well-defined and strong security posture, and which can hold their own in adversarial situations (\eg compliance assessments) within certain {\em known} bounds, which will eventually lead to long-lasting security benefits for end-user software.

\section*{Acknowledgment}
We thank the tool designers for being open to discussion, suggestions, and improvements.
Moreover, we thank the anonymous reviewers for constructive feedback.
We also thank Jake Zappin and Johnny Clapham, William \& Mary, for contribution in extending the taxonomy.
Finally, the authors have been supported in part by the NSF-$2237012$ and NSF-$2132281$ grants and Coastal Virginia Center for Cyber Innovation and the Commonwealth Cyber Initiative, an investment in
the advancement of cyber R\&D, innovation, and workforce development.
For more information about COVA CCI and CCI, visit www.covacci.org and www.cyberinitiative.org. Any opinions, findings and conclusions expressed herein are the authors' and do not necessarily reflect those of the sponsors.

\bibliographystyle{ACM-Reference-Format}
\bibliography{bib/bibliography,bib/onlinebibliography,bib/phone,bib/cryptoapi,bib/taint,bib/iot,bib/mutation-references,bib/misc,bib/mutation_security}

\appendix

\section{Appendix: Code Snippets}\label{app:code_snippets}

\begin{lstlisting}[frame=tb,caption={\small Predictable/Non-Random Derivation of Value (\opnumber{6})}, label={lst:bad_derivation_operator},language=java]
v=new Date(System.currentTimeMillis()).toString();new IvParameterSpec(v.getBytes(),0,8);}
\end{lstlisting}
\vspace{-1em}
\begin{lstlisting}[frame=tb,caption={{\small Exception in an {\em always-false} condition block (\opnumber{7}).}}, label={lst:conditional_exception},language=java]
void checkServerTrusted(X509Certificate[] x, String s)throws CertificateException {
if (!(null != s && s.equalsIgnoreCase("RSA"))) {throw new CertificateException("not RSA");}
\end{lstlisting}
\vspace{-1em}
\begin{lstlisting}[frame=tb,caption={\small False return within an {\em always true} condition block (\opnumber{8}).}, label={lst:condition_return},language=java]
public boolean verify(String host, SSLSession s) {if(true || s.getCipherSuite().length()>=0)}return true;} return false;}
\end{lstlisting}
\vspace{-1em}
\begin{lstlisting}[frame=tb,caption={\small Implementing an Interface with no overridden methods.}, label={lst:abstract_extension_empty},language=java]
interface ITM extends X509TrustManager { } abstract class ATM implements X509TrustManager { }
\end{lstlisting}
\vspace{-1em}
\begin{lstlisting}[frame=tb,caption={Inner class object from Abstract type ({\bf OP$_{12}$})}, label={lst:abstract_inner_class},language=java]
new HostnameVerifier(){public boolean verify(String h, SSLSession s) {return true; } };
\end{lstlisting}
\vspace{-1em}
\begin{lstlisting}[frame=tb,caption={\small Anonymous Inner Class Object of \texttt{X509ExtendedTrustManager} (\ref{flaw:X509ExtendedTrustManager})}, label={lst:aic_X509ExtendedTrustManager},language=java]
new X509ExtendedTrustManager(){
public void checkClientTrusted(X509Certificate[] chain,String a) throws CertificateException{}
public void checkServerTrusted(X509Certificate[] chain,String a)throws CertificateException{}
public X509Certificate[] getAcceptedIssuers() {return null;} ...};
\end{lstlisting}
\vspace{-1em}
\begin{lstlisting}[frame=tb,caption={\small Specific Condition in \inlinesmall{checkServerTrusted} method (\ref{flaw:X509TrustManagerSpecificConditions}) }, label={lst:specific_condition_trustmanager},language=java]
void checkServerTrusted(X509Certificate[] certs, String s) throws CertificateException{
 if (!(null != s || s.equalsIgnoreCase("RSA") || certs.length >= 314)) {
   throw new CertificateException("Error");}}
\end{lstlisting}
\vspace{-1em}
\begin{lstlisting}[frame=tb,caption={\small Anonymous Inner Class Object of An Empty Abstract Class that implements \hostnameVerifier}, label={lst:aic_empty_ext_abstract_hostname},language=java]
abstract class AHV implements HostnameVerifier{} new AHV(){
    public boolean verify(String h, SSLSession s) return true;};
\end{lstlisting}
\vspace{-1em}
\begin{lstlisting}[frame=tb,caption={\small Irrelevant Loop in \getAcceptedIssuers{} method of \xtrustManager}, label={lst:irrelevant_loop_trustmanager},language=java]
public X509Certificate[] getAcceptedIssuers() {
for(int i = 0; i<100; i++){if (i==50)return EMPTY_X509CERTIFICATE_ARRAY;}
return EMPTY_X509CERTIFICATE_ARRAY;}
\end{lstlisting}
\vspace{-1em}
\begin{lstlisting}[frame=tb,caption={Anonymous inner class object with a vulnerable \checkServerTrusted method}, label={lst:aic_x509tm},language=java]
abstract class AbstractTM implements X509TrustManager{} new AbstractTM(){
public void checkServerTrusted(X509Certificate[] c, String a) throws CertificateException {}
public X509Certificate[] getAcceptedIssuers() {return null;}};
\end{lstlisting}
\vspace{-1em}
\begin{lstlisting}[frame=tb,caption={\small Anonymous Inner Class Object of an Interface that extends \hostnameVerifier}, label={lst:aic_empty_ext_interface_hostname},language=java]
interface IHV extends HostnameVerifier{}
new IHV(){ public boolean verify(String h, SSLSession s) return true;};
    \end{lstlisting}
\vspace{-1em}
\begin{lstlisting}[frame=tb,caption={\small Misuse case requiring a trivial new operator}, label={lst:trivial},language=java]
KeyGenerator keyGen = KeyGenerator.getInstance("AES");
keyGen.init(128); SecretKey secretKey=keyGen.generateKey();
    \end{lstlisting}
\vspace{-1em}
\begin{lstlisting}[frame=tb,caption={\small \cryptoguard's code ignoring names with "android"}, label={lst:android-dot},language=java]
if (!className.contains("android."))
    classNames.add(className.substring(1, className.length() - 1)); return classNames;
 \end{lstlisting}
\vspace{-1em}
\begin{lstlisting}[frame=tb,caption={\small Generic Conditions in \checkServerTrusted{}}, label={lst:xtrustManagerGenericConditions},language=java]
if(!(true || arg0==null || arg1==null)) { throw new CertificateException();}
\end{lstlisting}
\vspace{-1em}
\begin{lstlisting}[frame=tb,caption={\small Transformation String formation in Apache Druid similar to \fnumber{2} which uses \AES in \texttt{CBC} mode with \texttt{PKCS5Padding}, a configuration that is known to be a misuse~\cite{FBX+17,owasp:mislabel}.}, label={lst:transformationstring:apache:druid},language=java]
this.name = name == null ? "AES" : name;
this.mode = mode == null ? "CBC" : mode;
this.pad = pad == null ? "PKCS5Padding" : pad;
this.string = StringUtils.format("%
\end{lstlisting}
\vspace{-1em}
\begin{lstlisting}[frame=tb,caption={\small Iterative Method Chaining ({\bf OP$_{13}$})}, label={lst:iterativechaining},language=java]
Class T {
  int i = 0; cipher = "AES/GCM/NoPadding";
  public void A(){cipher = "AES/GCM/NoPadding";}
  public void B(){cipher = "AES/GCM/NoPadding";}
  public void C(){cipher = "AES/GCM/NoPadding";}
  public void D(){cipher = "AES";} public String getVal(){return cipher;}}
Cipher.getInstance(new T().A().B().C().D().getVal() ) ;
\end{lstlisting}
\vspace{-1em}
\begin{lstlisting}[frame=tb,caption={\small Iterative Conditionals}, label={lst:iterativeconditionals},language=java]
Class T {
    int i = 0;cipher = "AES/GCM/NoPadding";
    public void A(){
    if (i == 0){if (i == 0){if(i == 0){cipher = "AES";}
            else{cipher = "AES/GCM/NoPadding";}}
        else{cipher = "AES/GCM/NoPadding";}
    } else{cipher = "AES/GCM/NoPadding";}}
    public String getVal(){return cipher;}
}Cipher.getInstance(new T().A().getVal() ) ;
\end{lstlisting}
\vspace{-1em}

\begin{lstlisting}[frame=tb,caption={\small Method Builder ({\bf OP$_{15}$}) }, label={lst:methodbuilder},language=java]
Class T {cipher = "AES/GCM/NoPadding";
    public String A(){return "D";}public String B(){return "E";}public String C(){return "S";}
    public void add(){cipher = A() + B() + C();} public String getVal(){return cipher;}}
Cipher.getInstance(new T().add().getVal() ) ;
\end{lstlisting}
\vspace{-1em}
\begin{lstlisting}[frame=tb,caption={\small Object Sensitive, using the object created in Listing A.1}, label={lst:ObjectSensitivity},language=java]
T sec = new T();T notsec = new T().mthd2();sec = notsec;Cipher.getInstance(sec.getVal());
\end{lstlisting}
\vspace{-1em}
\begin{lstlisting}[frame=tb,caption={\small Build Variable}, label={lst:buildvariable},language=java]
String var = "AES";char[] var1  = var.toCharArray();Cipher.getInstance(String.valueOf(var));
\end{lstlisting}
\vspace{-1em}
\begin{lstlisting}[frame=tb,caption={\small Substring}, label={lst:substring},language=java]
Cipher.getInstance("secureParamAES".substring(11));
\end{lstlisting}
\vspace{-1em}
\begin{lstlisting}[frame=tb,caption={\small Constant IV}, label={lst:constantiv},language=java]
byte[] cryptoTemp = "12345678".getBytes();
javax.crypto.spec.IvParameterSpec ivSpec = new javax.crypto.spec.IvParameterSpec.getInstance(cryptoTemp,"AES");
\end{lstlisting}

\end{document}